\documentclass[aps,pre,floats,superscriptaddress,nofootinbib,floatfix,twocolumn,longbibliography,showkeys]{revtex4-1}
\usepackage{amssymb,amsmath,amsfonts,amsthm}
\usepackage{graphicx}
\usepackage{newtxtext,newtxmath}
\usepackage{dcolumn}
\usepackage{bm}
\usepackage{comment}
\usepackage{csquotes}
\usepackage{mathtools}
\usepackage{gensymb}
\usepackage{hyperref}
\usepackage{makecell,orcidlink}
\usepackage{changepage}
\usepackage{multirow}
\usepackage[table]{xcolor}
\usepackage[mathscr]{euscript}
\usepackage{ulem}
\usepackage{comment}
\graphicspath{{./fig/}}

\hypersetup{
	colorlinks=true,
	linkcolor=blue,
	filecolor=blue,      
	urlcolor=blue,
	citecolor=blue,
}

\begin{document}

\title{Beyond the Markovian limit:\\ Exact solutions for active motion in a power-law viscoelastic bath}
\author{Mintu Karmakar\,\orcidlink{0000-0003-1037-2580}}
\email{mkarmakar094@ucas.ac.cn, mkarmakar094@ub.edu}
\affiliation{Wenzhou Institute of the University of Chinese Academy of Sciences, Wenzhou, Zhejiang 325011, China.}
\affiliation{School of Physical Sciences, University of Chinese Academy of Sciences, Beijing 100049, China.}
\affiliation{Departament de F\'isica de la Mat\`eria Condensada, Universitat de Barcelona, Mart\'i i Franqu\`es 1, E08028 Barcelona, Spain.}

\author{Jure Dobnikar\,\orcidlink{0000-0002-1169-6619}}
\email{jd489@cam.ac.uk}
\affiliation{Chinese Academy of Sciences Key Laboratory of Soft Matter Physics, Institute of Physics, Chinese Academy of Sciences, Beijing 100190, China}
\affiliation{School of Physical Sciences, University of Chinese Academy of Sciences, Beijing 100049, China.}
\affiliation{Wenzhou Institute of the University of Chinese Academy of Sciences, Wenzhou, Zhejiang 325011, China.}

\author{Ignacio Pagonabarraga\,\orcidlink{0000-0002-6187-5025}}
\email{ipagonabarraga@ub.edu}
\affiliation{Departament de F\'isica de la Mat\`eria Condensada, Universitat de Barcelona, Mart\'i i Franqu\`es 1, E08028 Barcelona, Spain.}
\affiliation{UBICS University of Barcelona Institute of Complex Systems , Mart\'i i Franqu\`es 1, E08028 Barcelona, Spain}
%\date{}

\begin{abstract}
Active particles from bacteria to synthetic microswimmers often navigate viscoelastic media with complex relaxation dynamics.
The classical active Brownian model that assumes instantaneous friction is clearly not applicable to describe such motility, while the non-Markovian processes combined with viscoelasticity are relatively unexplored. 
Here, we develop an analytical theory for an active particle in a power-law viscoelastic medium by solving coupled non-Markovian generalized Langevin equations for translational and rotational degrees of freedom. 
The viscoelastic memory results in novel phenomena such as fractional short-time transport, enhanced long-time persistence, and de-correlation of the instantaneous force and the swimmer orientation. We demonstrate that the memory kernel controls the anomalous scaling exponents, while the activity determines the crossover between sub-diffusive, ballistic and diffusive regimes. Our work provides a framework for theoretical description of biological and synthetic micro swimmers in complex biological and polymeric environments.

\end{abstract}

\maketitle

\section{Introduction}
\label{sec:introduction}
%The study of active matter has emerged as a central theme in non-equilibrium statistical physics, aiming to understand systems composed of self-propelled agents that convert stored or ambient energy into systematic motion~\cite{Bechinger2016, Marchetti2013, Ramaswamy2010, Gompper2020}. 
From bacterial swarms~\cite{Zhang2010, Ariel2015} and cellular monolayers~\cite{Angelini2011, Garcia2015} to synthetic microswimmers~\cite{Howse2007, Walther2013, Ebbens2016, Bechinger2016}, active systems exhibit emergent collective phenomena such as flocking, swarming, clustering, dynamic pattern formation and motility induced phase separation (MIPS)~\cite{vicsek1995novel, Vicsek2012, gregoire2004onset, chate2008modeling, chate2008collective, Bricard2013, Palacci2013, Theeyancheri2024, Codina2022}. At the single-particle level, the canonical Active Brownian Particle (ABP) model~\cite{Romanczuk2012, Fily2012}, assumes particles with a constant self-propulsion speed subject to rotational diffusion. 
This minimal framework explains a wide range of nonequilibrium behaviors in dilute and interacting systems~\cite{Solon2015, Stenhammar2014, Redner2013, Farage2015, Berthier2014, Bialke2013, Turci2021wetting, Marchetti2013, Ramaswamy2010, Gompper2020}.\\ 

A central assumption of the ABP model is that the surrounding fluid is Newtonian and therefore the action of friction is instantaneous and memoryless.
In contrast, many biological and technological environments are complex fluids characterized by viscoelasticity and long-time memory. 
Examples include cytoplasm~\cite{Weigel2011, Tabei2013}, mucus~\cite{Chen2007, Cone2004}, polymer solutions~\cite{Patteson2015, Zottl2019}, and dense colloidal suspensions~\cite{Gomez-Solano2016, Narinder2018}. 
In these media, the mechanical response depends on the history of deformation and cannot be described by instantaneous friction alone~\cite{Larson1999, Doi1988}. 
%Understanding active motion in such environments requires a theoretical framework incorporating temporal memory.
The Generalized Langevin Equation (GLE) provides a theoretical framework to incorporate memory effects through a time-dependent friction kernel~\cite{Kubo1966, Zwanzig2001, Coffey2012}. 
The choice of the kernel reveals the complexity of the relaxation behaviour; simple analytical approaches such as the Maxwell model~\cite{Sprenger2022, Sevilla2019, Ghosh2015, ouyang2023swimming} assume an exponentially decaying kernel with a single relaxation timescale, while a kernel with a broad relaxation spectra~\cite{Mason1995, Waigh2005} is essential to capture the power-law stress relaxation, persistent anomalous diffusion, and long-range temporal correlations, which are hallmarks of complex soft materials such as
polymeric networks, biological gels, and glassy systems~\cite{Larson1999, Hofling2013, Sollich1997, Bouchaud1990}. 
The active Ornstein-Uhlenbeck model (AOUM) incorporating a broad family of memory kernels into the self-propulsion dynamics provides analytical and numerical results for power-law decaying memory in one-dimensional active particle systems~\cite{Sevilla2019}, but a systematic analytical treatment of active Brownian motion with long-range power-law memory in two- or three-dimensional media is currently still lacking.
Power-law friction kernels provide a promising model to describe microswimmer locomotion in non-Newtonian environments. The power-law form of the kernel is physically motivated by the fractional nature of dynamics observed in a wide range of soft matter systems~\cite{Metzler2000, Barkai2012, Metzler2014}, including those with fractal-like polymer networks~\cite{Granick1995soft}, crowded intracellular environments~\cite{Guigas2007, Weber2012nonthermal}, and glassy dynamics~\cite{Sollich1998}. 
The algebraic decay of such kernels introduces long-range temporal correlations altering the particle's dynamics into a non-Markovian process~\cite{Hanggi1990, Goychuk2012}. This leads to a rich phenomenology governed by fractional calculus~\cite{Mainardi2010, Sandev2017}, where familiar concepts like diffusion and relaxation are replaced by their anomalous counterparts~\cite{Lutz2001, Burov2010, Gomez2020active}. 

In this work, we develop an analytical theory for a two-dimensional ABP in a viscoelastic medium characterized by a power-law memory kernel. 
We formulate coupled non-Markovian generalized Langevin equations for both translational and rotational degrees of freedom, thereby explicitly accounting for the interplay between translational friction memory and rotational persistence. 
In contrast with the active Ornstein-Uhlenbeck formulations where memory acts directly on the propulsion velocity~\cite{Sevilla2019}, we retain a constant self-propulsion speed and allow viscoelasticity to act on the translational and rotational dynamics separately.
By analytically solving the model, we demonstrate that viscoelastic memory qualitatively reshapes active dynamics. 
Orientational correlations decay via stretched exponentials, short-time transport becomes fractional rather than diffusive, and a finite delay emerges between the propulsion force and particle orientation -- a phenomenon entirely absent in Newtonian active Brownian motion. 
%These results establish viscoelastic active matter as a distinct dynamical regime beyond classical Markovian descriptions and provide experimentally testable signatures for microswimmers in complex fluids.

The remainder of this paper is organized as follows. In Sec. \ref{sec:model}, we introduce the theoretical model, detailing the Generalized Langevin Equations with a power-law memory kernel. Sec. \ref{sec:results} constitutes the analytical solutions for the key observables. Finally, in Sec. \ref{sec:conclusion}, we summarize our principal findings and discuss their implications and promising directions for future research.

\section{The Model: Active Motion with Power-Law Memory}
\label{sec:model}
We model a single, self-propelled particle, representing an active colloid or microswimmer, moving in a two-dimensional, homogeneous, and isotropic viscoelastic medium at a finite temperature $T$~\cite{tenHagen2011, Lowen2020}. The particle's state at time $t$ is fully described by its position vector $\mathbf{r}(t) = (x(t), y(t))$ and its orientation angle $\phi(t)$. The orientation is represented by a unit vector $\hat{n}(t) = (\cos\phi(t), \sin\phi(t))$. The particle moves with a constant self-propulsion speed $v_0$ along its orientation vector $\hat{n}(t)$. Additionally, we allow for an intrinsic tendency to perform circular motion, characterized by a constant angular velocity $\omega_0$. We consider the overdamped limit, where inertial effects are negligible, which is appropriate for motion on the colloidal scale.

The dynamics of the particle are governed by a set of coupled, non-Markovian Generalized Langevin Equations (GLEs) for the translational and rotational degrees of freedom~\cite{Zwanzig1973, Fox1977, Adelman1976}. These equations capture the time-delayed response of the viscoelastic medium through a memory kernel. Assuming the process starts at $t=0$ from an initial state $(\mathbf{r}_0, \phi_0)$, the equations are given by:
\begin{align}
\int_{0}^{t} \Gamma_T(t-t') \left( \dot{\mathbf{r}}(t') - v_0 \hat{n}(t') \right) dt' &= \boldsymbol{\xi}(t) \tag{1a} \label{eq:translational} \\
\int_{0}^{t} \Gamma_R(t-t') \left( \dot{\phi}(t') - \omega_0 \right) dt' &= \eta(t) \tag{1b} \label{eq:rotational}
\end{align}
Here, $\Gamma_T(t)$ and $\Gamma_R(t)$ are the translational and rotational memory kernels, respectively. The integrals represent the retarded frictional forces and torques exerted by the fluid. The terms $v_0 \hat{n}(t')$ and $\omega_0$ represent the active velocity and intrinsic angular velocity. Physically, $v_0\hat n(t')$ denotes the intrinsic, force-free swimming velocity that the propulsion mechanism maintains relative to the local medium, and $\omega_0$ the corresponding intrinsic angular velocity. The viscoelastic bath opposes departures of the instantaneous particle velocity (angular velocity) from these preferred values, so the retarded friction acts on the slip $\dot{\mathbf r}-v_0\hat n$ ($\dot\phi-\omega_0$) rather than on $\dot{\mathbf r}$ ($\dot\phi$) alone. In the memoryless limit $\Gamma_T(t)\to \gamma_T\delta(t)$, Eq.~(\ref{eq:translational}) reduces to $\gamma_T(\dot{\mathbf r}-v_0\hat n)=\boldsymbol\xi$, i.e.\ $\dot{\mathbf r}=v_0\hat n+\boldsymbol\xi/\gamma_T$, recovering the conventional active drift velocity $v_0\hat n$. The same structure, adopted previously for viscoelastic active particles~\cite{Sprenger2022}, preserves the equilibrium fluctuation-dissipation relation between noise and friction.  $\boldsymbol{\xi}(t)$ and $\eta(t)$ are the stochastic force and torque, respectively, which arise from thermal fluctuations. We define a circling swimmer as a particle with nonzero intrinsic angular velocity $\omega_0$, which undergoes deterministic rotation even in the absence of noise. A non-circling swimmer corresponds to $\omega_0=0$, where rotational dynamics arises solely from stochastic fluctuations.

The central feature of our model is the choice of a power-law memory kernel, which accounts for the long-range temporal correlations characteristic of complex fluids with a broad spectrum of relaxation modes~\cite{Goychuk2012, Lutz2001}. For both translational and rotational motion, the kernel takes the form:
\begin{equation}
\Gamma_{k}(t) = \frac{c_{k}}{\Gamma(1-\alpha_{k})} t^{-\alpha_{k}} \quad \text{for } t > 0 \tag{2} \label{eq:memory_kernel}
\end{equation}
where the subscript $k$ stands for either $T$ (translational) or $R$ (rotational). The parameter $c_k\equiv c_{\alpha_k}$ is the fractional friction strength. It carries exponent-dependent units, and $\alpha_k$ is the memory exponent, constrained to the range $0 < \alpha_k < 1$~\cite{di2011visco, bonfanti2020fractional}. This form is motivated by fractional calculus and provides an economical two-parameter description of power-law rheology, and has been successfully used to model anomalous diffusion in various contexts~\cite{Kou2008, Evangelista2018fractional, Luchko2016new}. We stress that the pure power law is an idealization valid over an intermediate window of times (or frequencies): real viscoelastic media exhibit a lower (microscopic) and an upper (terminal-relaxation) cutoff, beyond which the algebraic response crosses over to simple-fluid behaviour. Our results apply within this intermediate scaling window. This exponent dictates the nature of the memory: in the limit $\alpha_k \to 1$, the kernel approaches a Dirac delta function, recovering the memoryless friction of a simple Newtonian fluid. As $\alpha_k$ decreases towards 0, the kernel's decay becomes slower, representing an increasingly persistent memory. The Newtonian limit is singular and must be understood distributionally: since $\Gamma(1-\alpha_k)\to\infty$ as $\alpha_k\to1^-$, the kernel amplitude vanishes for every $t>0$ while its total weight is conserved, so $\Gamma_k(t)\to c_k\,\delta(t)$ in the weak sense; equivalently, in Laplace space $\tilde\Gamma_k(s)=c_k s^{\alpha_k-1}\to c_k$, the transform of $c_k\delta(t)$. In this limit the fractional friction strength reduces to the ordinary friction coefficient ($c_T\to\gamma_T$, $c_R\to\gamma_R$), so that Eqs.~(\ref{eq:translational})--(\ref{eq:rotational}) recover the memoryless active Brownian particle.

The stochastic terms $\boldsymbol{\xi}(t)$ and $\eta(t)$ are modeled as zero-mean, stationary Gaussian processes. We assume the system is in thermal equilibrium in the absence of activity, meaning the noise and dissipation originate from the same underlying microscopic interactions. Consequently, their correlation functions are determined by the memory kernels via the second Fluctuation-Dissipation Theorem (FDT)~\cite{Kubo1966, Marconi2008}, which connects the magnitude of thermal fluctuations to the dissipative frictional forces in equilibrium:
\begin{align}
\langle \xi_i(t) \xi_j(t') \rangle &= k_B T \Gamma_T(|t-t'|) \delta_{ij} \tag{3a} \label{eq:fdt_trans} \\
\langle \eta(t) \eta(t') \rangle &= k_B T \Gamma_R(|t-t'|) \tag{3b} \label{eq:fdt_rot}
\end{align}
where $k_B$ is the Boltzmann constant, $T$ is the absolute temperature, and $\delta_{ij}$ is the Kronecker delta. Furthermore, the translational and rotational noises are assumed to be uncorrelated, $\langle \xi_i(t) \eta(t') \rangle = 0$, reflecting the distinct degrees of freedom. This set of equations provides a complete theoretical framework for investigating the dynamics of an active particle in a fluid with power-law viscoelasticity.

Equations~(\ref{eq:translational})--(\ref{eq:rotational}) are adopted as a phenomenological non-Markovian model in which the active drive biases an otherwise equilibrium viscoelastic bath, so that the noise and friction obey the equilibrium second FDT, Eqs.~(\ref{eq:fdt_trans})--(\ref{eq:fdt_rot}). For a Gaussian bath this structure can be justified microscopically by exact projection of a time-dependent many-body Hamiltonian carrying a non-equilibrium driving force~\cite{Netz2024}, which reproduces the form of Eq.~(\ref{eq:translational}); for non-Gaussian baths the resulting generalized Langevin equation and the associated fluctuation relation take a different form, so Eqs.~(\ref{eq:translational})--(\ref{eq:fdt_rot}) should be understood as the Gaussian-bath case.

The analytical tractability of the model rests on three assumptions. (i)~The GLEs are linear in $\dot{\mathbf r}$ and $\dot\phi$, so they can be solved by Laplace transform. (ii)~The noises are Gaussian and satisfy the FDT, which makes the angular displacement $\Delta\phi_{\rm stoch}(t)$ a zero-mean Gaussian variable; the orientational correlation then follows exactly from the characteristic function $\langle e^{i\Delta\phi_{\rm stoch}}\rangle=e^{-\frac12\langle\Delta\phi_{\rm stoch}^2\rangle}$. (iii)~The self-propulsion speed $v_0$ is constant, so the active-velocity correlation reduces to $v_0^2 C(t)$ and the mean-square displacement becomes a single quadrature over $C(t)$. Relaxing any of these---a fluctuating speed, non-Gaussian noise, or cross-correlated translational and rotational noise---would break the Gaussian reduction and, in general, require a perturbative or numerical treatment; the present closed forms then serve as the natural reference limit. We also note that the analytical method itself applies to any memory kernel with a well-defined Laplace transform: the stretched-exponential orientational relaxation and the fractional short-time scaling are specific to the power-law kernel $\tilde\Gamma_k(s)\propto s^{\alpha_k-1}$, whereas the emergence of a force--orientation delay and the sub-ballistic/ballistic/diffusive crossover sequence are generic consequences of temporal nonlocality. An independent and concurrent analytical study of active Brownian motion in power-law viscoelastic media reaches consistent conclusions~\cite{Quevedo2025}.

\section{Results and Discussion}
\label{sec:results}

The system is characterized by two memory exponents ($\alpha_T, \alpha_R$), which are dimensionless parameters (ranging from 0 to 1)  governing the "fractional" nature of translational and rotational memory, respectively. A value of 1 corresponds to the memoryless Newtonian case. 
We introduce the dimensionless time $\tilde{t} = t / \tau_R$ where $\tau_R$ is the characteristic rotational fractional memory time scale: $\tau_R = \left(\frac{c_R}{k_BT}\right)^{1/\alpha_R}$. 
The dimensionless rotational frequency ($\tilde{\omega}_0$) is consequently $\tilde{\omega}_0 = \omega_0 \tau_R$. 
Lengths are measured in units of the active persistence length $L_p=v_0\tau_R$, and the strength of activity is quantified by the P\'eclet number
\begin{equation}
\mathrm{Pe}^2\equiv\frac{v_0^2\,c_T\,\tau_R^{\,2-\alpha_T}}{k_BT}=\left(\frac{L_p}{\ell_{\rm th}}\right)^2,
\tag{3c}
\label{eq:peclet_def}
\end{equation}
the ratio of $L_p$ to the thermal (fractional) length $\ell_{\rm th}=(k_BT\,\tau_R^{\alpha_T}/c_T)^{1/2}$ explored by translational memory over one rotational time. This is the fractional generalization of the P\'eclet numbers used in active matter: in the Markovian limit $\alpha_T,\alpha_R\to1$ one has $c_T\to\gamma_T$, $c_R\to\gamma_R$ and $\tau_R\to1/D_R$, so that $\mathrm{Pe}^2\to v_0^2/(D_TD_R)$. Writing the conventional active and thermal P\'eclet numbers as $\mathrm{Pe}_{\rm act}=v_0/(\sigma D_R)$ and $\mathrm{Pe}_{\rm th}=v_0\sigma/D_T$, this is their geometric mean, $\mathrm{Pe}\to\sqrt{\mathrm{Pe}_{\rm act}\,\mathrm{Pe}_{\rm th}}$, in which the particle size $\sigma$ cancels---consistent with the fact that the present model contains no particle size. Since $L_p$ is an active scale, it is the appropriate unit in the active regime studied here; in the passive limit $v_0\to0$ one would instead rescale lengths by $\ell_{\rm th}$, under which all dimensionless observables remain finite.
All subsequent results are presented within this dimensionless framework, enabling a universal interpretation of the particle dynamics.

\subsection{Orientation correlation function}
\label{orrientational_correlation}

The orientational correlation function 
\(
C(\tilde t)=\langle \hat n(\tilde t)\cdot \hat n(0)\rangle
\)
measures the micro-swimmer persistence. 
We calculate the exact form of $C(\tilde t)$ and analyze how the memory exponent $\alpha_R$ and the intrinsic rotation frequency $\tilde{\omega}_0$ independently and cooperatively reshape orientational relaxation.

The solution (derived in Appendix~\ref{sec:correlation_derivation_appendix})  is:
\begin{equation}
C(\tilde{t}) 
= \cos(\tilde{\omega}_0 \tilde{t}) 
\exp\!\left[-\frac{\tilde{t}^{\alpha_R}}{\Gamma(1+\alpha_R)}\right].
\tag{4}
\label{eq:orientation}
\end{equation}

Equation~(\ref{eq:orientation}) 
encodes the viscoelastic memory in the stretched exponential 
\(
\exp[-\tilde t^{\alpha_R}/\Gamma(1+\alpha_R)]
\),
and the intrinsic angular velocity in the oscillatory factor 
\(
\cos(\tilde{\omega}_0 \tilde t).
\)
We note that the rotational memory modifies the decay envelope of persistence without altering the intrinsic oscillation frequency. In a Newtonian (Markovian) fluid, corresponding to $\alpha_R=1$, Eq.~(\ref{eq:orientation}) reduces to the standard active Brownian result $C(\tilde t)=\cos(\tilde{\omega}_0 \tilde t)\,e^{-\tilde t}$ where the orientational relaxation is exponential and characterized by a single persistence time.
\begin{figure}[t]
\centering
\includegraphics[width=\columnwidth]{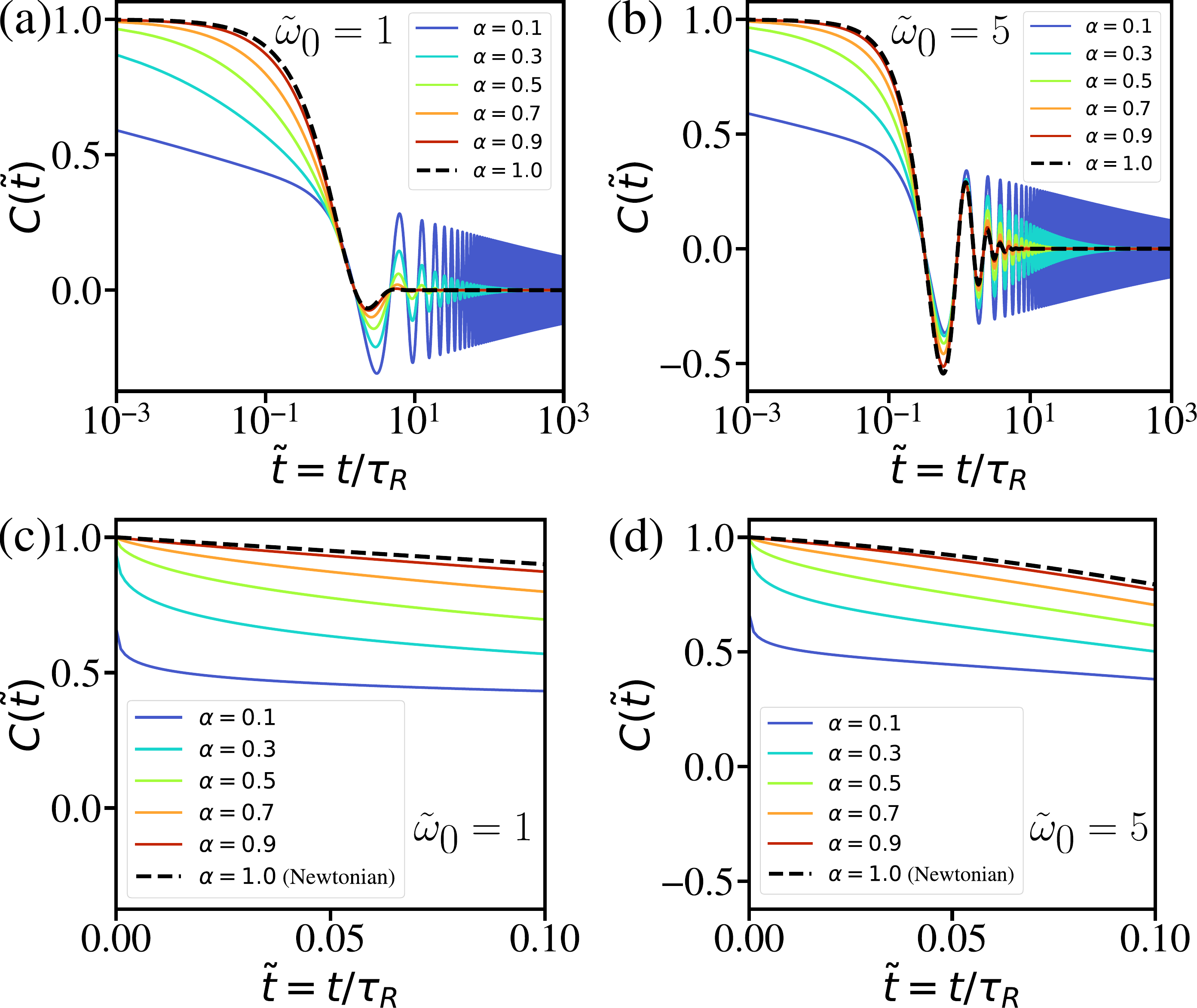}
\caption{(Color online) Orientational correlation function 
$C(\tilde t)=\langle \hat n(\tilde t)\cdot\hat n(0)\rangle$ 
as a function of dimensionless time $\tilde t=t/\tau_R$ for different memory exponents 
$\alpha=\alpha_R=\alpha_T$. 
(a) Moderate intrinsic angular velocity $\tilde{\omega}_0=1$. 
(b) Strong intrinsic angular velocity $\tilde{\omega}_0=5$. 
The dashed black curve corresponds to the Newtonian limit ($\alpha=1$), which exhibits exponential decay. 
For $\alpha<1$, the decay becomes stretched exponential, producing long-lived orientational persistence. 
Increasing memory (smaller $\alpha$) enhances the lifetime of oscillatory correlations without altering the intrinsic oscillation frequency. (c,d) Corresponding short-time behaviour of $C(\tilde t)$ in a linear scale exhibits rapid decay from 1 for smaller $\alpha$.}
\label{fig:orientational_correlation}
\end{figure}
\begin{figure*}[t]
\centering
\includegraphics[width=\textwidth]{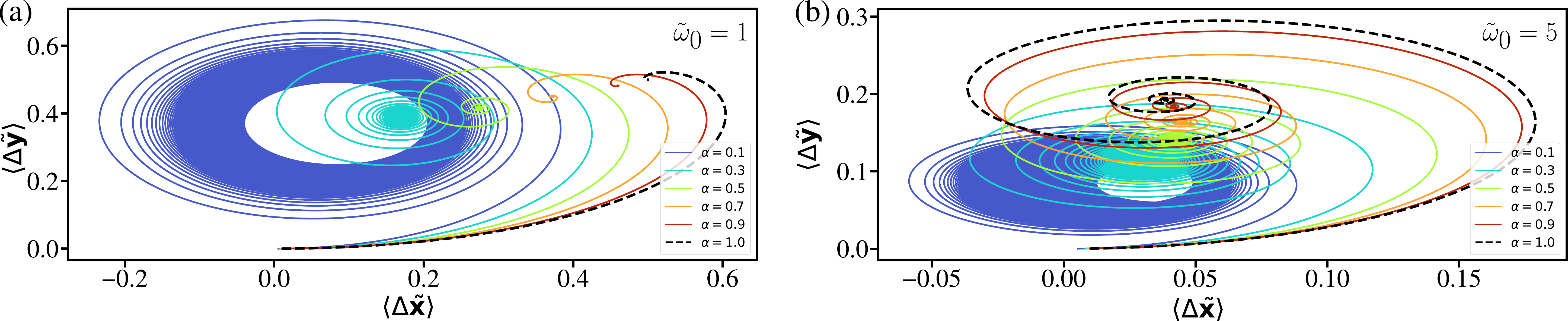}
\caption{(Color online) Mean trajectory of a circle swimmer in the $(\langle \Delta\tilde{\mathbf{x}}(\tilde t) \rangle ,\langle \Delta\tilde{\mathbf{y}} (\tilde t) \rangle)$ plane for different memory exponents $\alpha=\alpha_R=\alpha_T$. 
(a) $\tilde{\omega}_0=1$. 
(b) $\tilde{\omega}_0=5$. 
The dashed black curve corresponds to the Newtonian case ($\alpha=1$). Stronger viscoelastic memory (smaller $\alpha$) enhances orientational coherence and enlarges spiral excursions.}
\label{fig:mean_trajectory}
\end{figure*}
For $0<\alpha_R<1$, the decay envelope has a stretched exponential form, $C(\tilde t)\sim \exp(-\tilde t^{\alpha_R})$,
which is significantly slower and features a pronounced heavy tail at longer times (see Fig.~\ref{fig:orientational_correlation}(a-b)). 
At small $\alpha_R$, the medium effectively suppresses rapid randomization of orientation, thereby enhancing persistence over extended time intervals. 

At very short times ($\tilde t\ll 1$), the expansion of the Equation~(\ref{eq:orientation}) is:
\begin{equation}
   C(\tilde t)\approx 1-\frac{\tilde t^{\alpha_R}}{\Gamma(1+\alpha_R)}-\frac{\tilde{\omega}_0^2}{2}\tilde t^2+\cdots.
   \tag{5}
\label{eq:ini_Ct}
\end{equation}
In this regime memory modifies the initial decay of $C(\tilde t)$ through the fractional term $\tilde t^{\alpha_R}$. Importantly, for $\tilde t<1$ and $\alpha_R<1$, one has $\tilde t^{\alpha_R}>\tilde t$, so the initial reduction of $C(\tilde t)$ is actually stronger than in the Newtonian case 
Thus, although a smaller $\alpha_R$ produces a much slower decay at long times, it suppresses orientational coherence more strongly at  early times. 
This fractional short-time scaling is a direct fingerprint of power-law memory and can not be resolved by a single relaxation time exponential kernel previously studied~\cite{Sprenger2022}. 
The role of intrinsic angular velocity becomes particularly transparent when comparing Fig.~\ref{fig:orientational_correlation} (a) and (b). 
For moderate rotation ($\tilde{\omega}_0=1$, Fig.~\ref{fig:orientational_correlation}(a)), the memory-induced heavy tail primarily extends the persistence envelope. For stronger intrinsic rotation ($\tilde{\omega}_0=5$, Fig.~\ref{fig:orientational_correlation}(b)), however, the oscillatory nature of $C(\tilde t)$ becomes more pronounced, signifying that memory does not change the oscillation frequency but slows the decay of successive oscillations.

\subsection{Mean displacement}
Having established how memory modifies the orientational persistence, we explore how this altered persistence reshapes the mean trajectories of the swimmers. 

The mean displacement is obtained by integrating the mean orientation vector, as derived in Appendix~\ref{sec:mean_displacement_appendix}. We emphasize that the mean orientation vector decays with the envelope
$E(\tilde t)=\exp[-\tilde t^{\alpha_R}/\Gamma(1+\alpha_R)]$, whereas the scalar
correlation $C(\tilde t)=\cos(\tilde\omega_0\tilde t)E(\tilde t)$ additionally
carries the oscillatory factor; only the former enters the mean displacement. Accordingly,
\begin{equation}
\langle \Delta\tilde{\mathbf r}(\tilde t) \rangle 
=
\int_0^{\tilde t} 
E(\tilde t')
\left(
\cos(\phi_0+\tilde{\omega}_0 \tilde t')\,\hat{i}
+
\sin(\phi_0+\tilde{\omega}_0 \tilde t')\,\hat{j}
\right)
d\tilde t'.
\tag{6}
\label{eq:mean_displacement}
\end{equation}
The mean trajectory is determined by the oscillatory rotation imposed by $\tilde{\omega}_0$, while its envelope is controlled by the memory-modified decay factor $E(\tilde t)$. 
Figure~\ref{fig:mean_trajectory}(a,b) shows the mean trajectories for different $\alpha$ values at two different intrinsic frequencies $\tilde{\omega}_0$. 
Expanding the envelope $E(\tilde t')\approx 1-\tilde t'^{\alpha_R}/\Gamma(1+\alpha_R)$ together with the rotation factors in Eq.~(\ref{eq:mean_displacement}) for $\tilde t\ll 1$ at $\phi_0=0$, we obtain
\begin{align}
\langle \Delta\tilde x(\tilde t)\rangle
&\approx
\tilde t
-\frac{\tilde t^{1+\alpha_R}}{(1+\alpha_R)\Gamma(1+\alpha_R)}
-\frac{\tilde\omega_0^2}{6}\tilde t^3+\cdots,
\tag{7a}
\label{eq:short_x_memory}
\\
\langle \Delta\tilde y(\tilde t)\rangle
&\approx
\frac{\tilde\omega_0}{2}\tilde t^2
-\frac{\tilde\omega_0}{(2+\alpha_R)\Gamma(1+\alpha_R)}\tilde t^{2+\alpha_R}
+\cdots.
\tag{7b}
\label{eq:short_y_memory}
\end{align}
The terms $\langle \Delta\tilde x\rangle\sim \tilde t$ and $\langle \Delta\tilde y\rangle\sim (\tilde\omega_0/2)\tilde t^2$  are the leading terms at small times in the Newtonian case.
In this regime (dashed black lines in Figs.~\ref{fig:mean_trajectory}), the mean displacements initially follow a circular arc-like orbit whose characteristics are governed by the frequency $\tilde{\omega}_0$, and quickly saturate due to the exponential decay of correlations.
The effect of the memory is two-fold: since the memory-dependent short-time corrections are negative, decreasing $\alpha_R$ reduces both $\langle \Delta\tilde x\rangle$ and $\langle \Delta\tilde y\rangle$ -- resulting in an initial suppression of the arc at short times. 
At longer times, due to the slower stretched-exponential decay of $C(\tilde t)$, large spiral excursions are observed before the displacements saturate ({\sl e.g.}, green and blue lines in Fig.~\ref{fig:mean_trajectory}) . 
The trajectories at larger intrinsic angular velocity ($\tilde{\omega}_0=5$, Fig.~\ref{fig:mean_trajectory}(b)), are tightly wound, while at the lower value ($\tilde{\omega}_0=1$, Fig.~\ref{fig:mean_trajectory}(a)) they are more open.
The non-circular swimmers ($\tilde{\omega}_0=0$) exhibit long-lived orientationally persistent trajectories in a single direction (the initial direction) without the oscillatory component.
For completeness, we provide the orientational correlation and mean displacement plot for a non-circling swimmers in the Appendix~\ref{sec:appendix_noncircling}. 

\subsection{Mean-square displacement}
\label{msd_results}
\begin{figure}[t]
\centering
\includegraphics[width=\columnwidth]{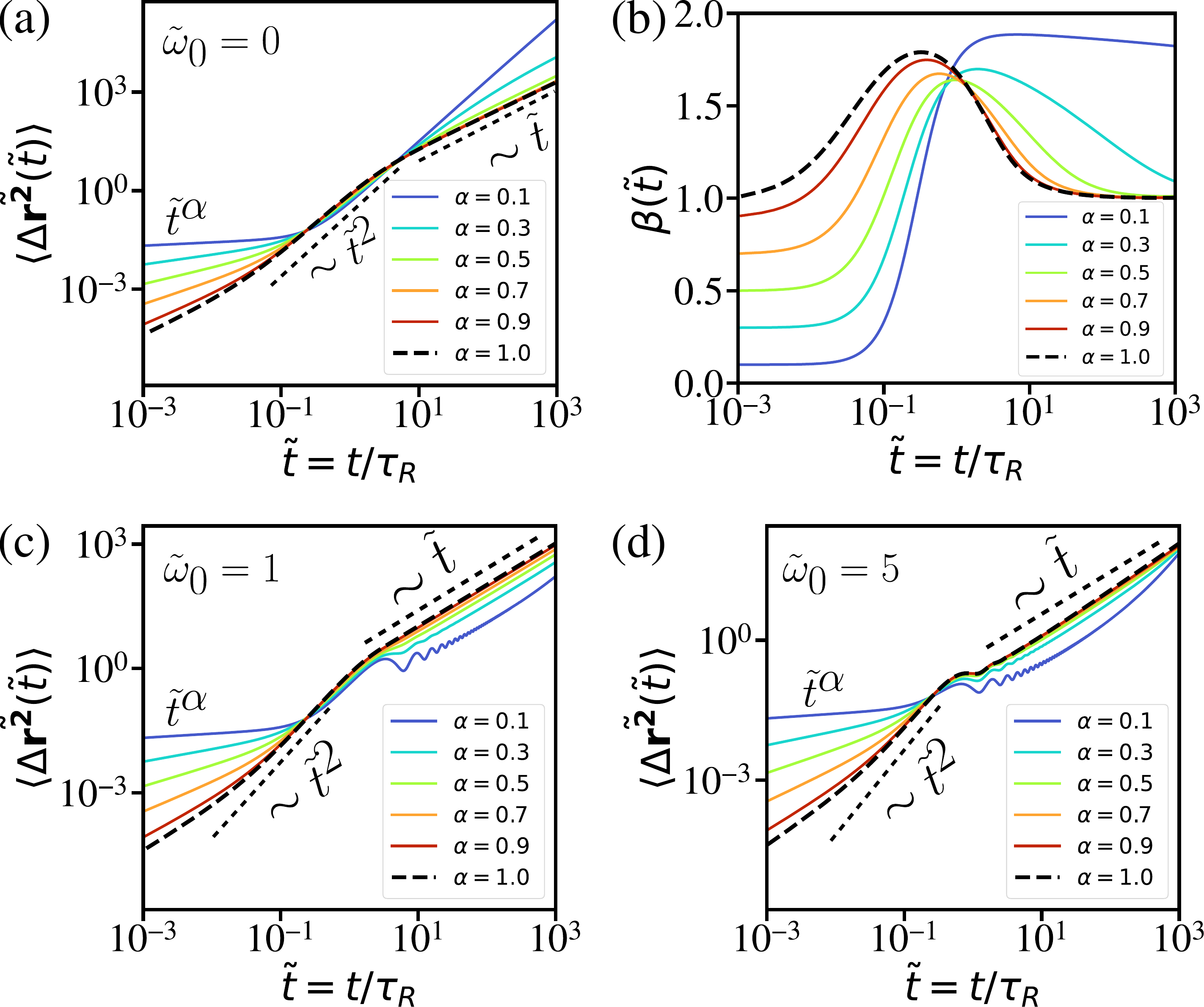}
\caption{(Color online) (a,b) Mean-square displacement 
$\langle\Delta\tilde{\mathbf{r^2}}(\tilde t)\rangle$ 
and the corresponding dynamical exponent $\beta(\tilde t)$ 
for different memory exponents $\alpha=\alpha_R=\alpha_T$ for non-circling swimmer ($\tilde{\omega}_0=0$). $\langle\Delta\tilde{\mathbf{r^2}}(\tilde t)\rangle$ for circling swimmer
(c) $\tilde{\omega}_0=1$. 
(d) $\tilde{\omega}_0=5$. 
The dashed black curves denote the Newtonian limit ($\alpha=1$). $\mathrm{Pe}=10$. 
Translational memory produces fractional short-time scaling ($\sim \tilde t^{\alpha_T}$), 
while rotational memory prolongs ballistic motion and enhances oscillatory transport signatures for chiral swimmers.}
\label{fig:msd}
\end{figure}

The mean displacement characterizes the average trajectory and its rotational asymmetries. 
The more convenient measurable quantifying the total exploratory dynamics including the effects of translational and rotational memory, and intrinsic angular velocity, is the mean-square displacement (MSD). 
As derived in Appendix~\ref{sec:msd_appendix}, the dimensionless MSD is
\begin{equation}
\langle\Delta\tilde{\mathbf{r^2}}(\tilde t)\rangle
=
\frac{4\,\tilde t^{\alpha_T}}{\mathrm{Pe}^2 \Gamma(1+\alpha_T)}
+
2 \int_0^{\tilde t}
(\tilde t-\tilde \tau)\,
C(\tilde \tau)
\, d\tilde \tau,
\tag{8}
\label{eq:msd_total}
\end{equation}
where the first term arises from translational memory, with $\mathrm{Pe}$ a dimensionless parameter measuring activity strength, and the second from active propulsion mediated by rotational correlations. Equation~(\ref{eq:msd_total}) reveals translational and rotational memory enter additively but affect different scaling regimes. At very short times, the MSD admits the expansion (see Appendix~\ref{asymp_MSD})
\[
\langle\Delta\tilde{\mathbf{r^2}}(\tilde t)\rangle
\sim
\frac{4}{\mathrm{Pe}^2 \Gamma(1+\alpha_T)}\,\tilde t^{\alpha_T}
+
\tilde t^2
+\cdots.
\]
Figure~\ref{fig:msd} shows the MSD 
$\langle\Delta\tilde{\mathbf r}^2(\tilde t)\rangle$ and the corresponding
dynamical exponent
\[
\beta(\tilde t)=\frac{d\log \langle\Delta\tilde{\mathbf r}^2(\tilde t)\rangle}
{d\log \tilde t}
\]
for non-circling swimmer ($\tilde{\omega}_0=0$, Fig.~\ref{fig:msd} (a,b)), and the MSD for circling swimmers with two different angular frequencies: $\tilde{\omega}_0=1$ (Fig.~\ref{fig:msd} (c)), and $\tilde{\omega}_0=5$ (Fig.~\ref{fig:msd} (a,b)).
The non-circling swimmer exhibits three transport regimes. 
At short times, the MSD follows a fractional scaling $\langle\Delta\tilde{\mathbf r}^2\rangle \sim \tilde t^{\alpha_T}$,
governed predominantly by the translational power-law memory. 
The first crossover occurs when the translational-memory term balances the
ballistic contribution from active propulsion, yielding the estimate crossover time (see Appendix~\ref{asymp_MSD})
\[
\tilde t_{c1}\sim
\left[\frac{4}{\mathrm{Pe}^2\Gamma(1+\alpha_T)}\right]^{1/(2-\alpha_T)}.
\]
Beyond this time the MSD enters the universal ballistic regime $\langle\Delta\tilde{\mathbf r}^2\rangle\sim \tilde t^2$ due to persistent self-propulsion. At longer times, orientational correlations decay and the motion crosses over to effective diffusion, $\langle\Delta\tilde{\mathbf r}^2\rangle \sim \tilde t$. This second crossover occurs at $\tilde t_{c2}\sim 2D_{\rm act}$, where $D_{\rm act}=\int_0^\infty C(\tau)\,d\tau$ is the active diffusivity (see Appendix~\ref{asymp_MSD}). These regimes are clearly reflected in the dynamical exponent $\beta(\tilde t)$ shown in Fig.~\ref{fig:msd}(b), which evolves from $\beta\simeq\alpha_T$ (fractional regime) to $\beta\simeq2$ (ballistic) and finally to $\beta\simeq1$ (diffusive). Figures~\ref{fig:msd}(c) and (d) show the MSD for circling swimmers ($\tilde{\omega}_0=1$ and $5$). Intrinsic rotation suppresses long-time transport in the Newtonian case ($\alpha=1$) due to rapid orientational dephasing. In viscoelastic media ($0<\alpha<1$), however, rotational memory slows this decorrelation through the stretched-exponential form of $C(\tilde t)$, thereby generating oscillatory intermediate-time transport. These oscillations become more pronounced for larger $\tilde{\omega}_0$ and smaller $\alpha$, reflecting long-lived rotational
coherence in the viscoelastic bath. Although the main figures set $\alpha_R=\alpha_T=\alpha$ for visual clarity, the two exponents act on distinct regimes and the formalism keeps them independent throughout. To make this explicit, Appendix~\ref{sec:appendix_alphaTR} shows the MSD and dynamical exponent for representative unequal pairs $(\alpha_T,\alpha_R)$: the short-time subdiffusive slope tracks the translational exponent $\alpha_T$ [so that $\beta(\tilde t)\to\alpha_T$ as $\tilde t\to0$], while the extent of the intermediate ballistic regime and the long-time crossover are governed by the rotational exponent $\alpha_R$.

\subsection{Effect of activity strength on anomalous transport}
\begin{figure}[t]
\centering
\includegraphics[width=\columnwidth]{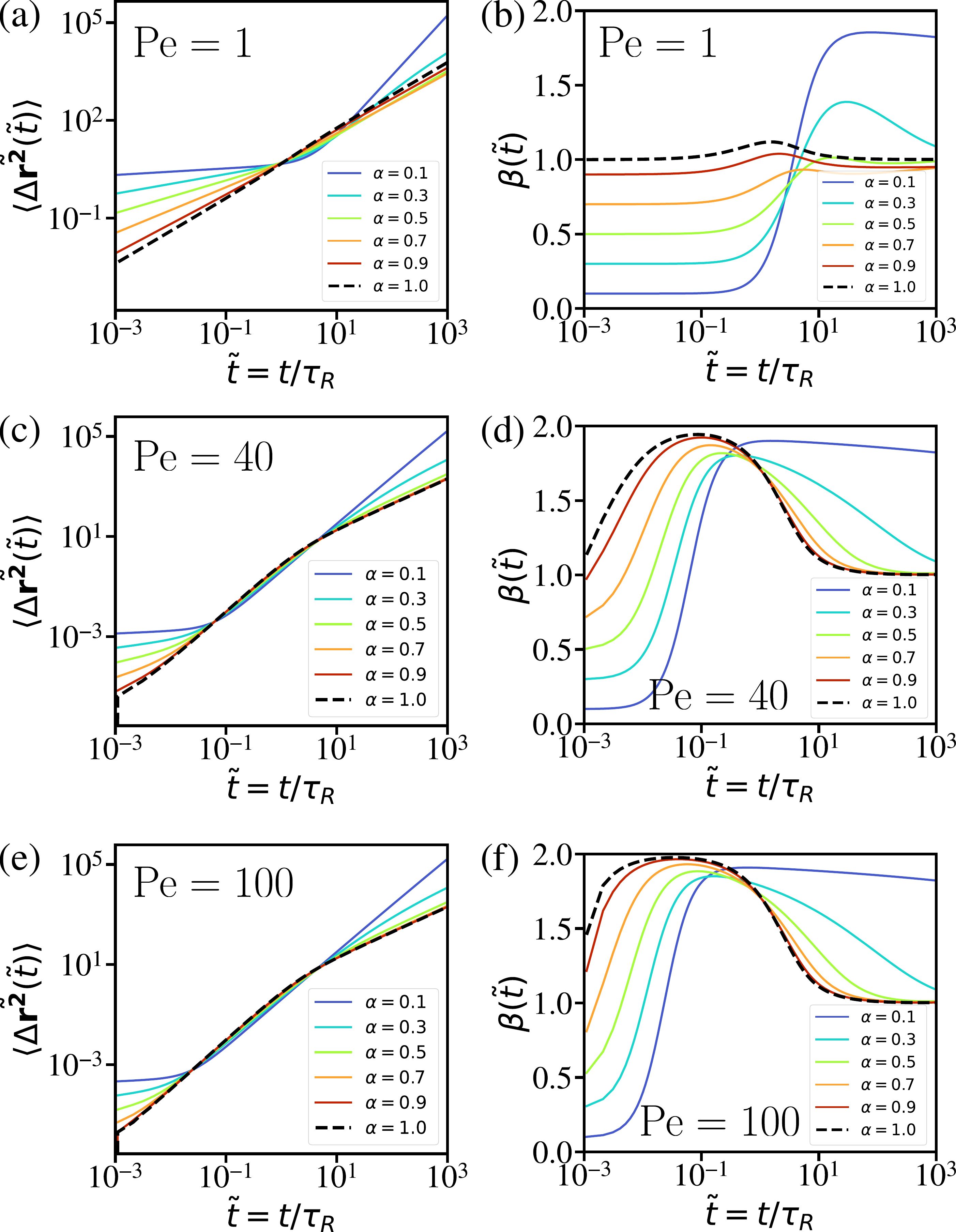}
\caption{(Color online) Influence of activity strength, $\mathrm{Pe}$, on the mean-square displacement 
$\langle\Delta\tilde{\mathbf{r^2}}(\tilde t)\rangle$ and dynamical exponent 
$\beta(\tilde t)$ 
for a non-circling swimmer ($\tilde{\omega}_0=0$). 
(a,b) $\mathrm{Pe}=1$, 
(c,d) $\mathrm{Pe}=40$, 
(e,f) $\mathrm{Pe}=100$. 
Different colors correspond to memory exponents $\alpha=\alpha_R=\alpha_T$, and the dashed black curve denotes the Newtonian limit ($\alpha=1$). 
Increasing $\mathrm{Pe}$ suppresses the relative contribution of the fractional thermal term and enhances the ballistic regime, while memory-induced anomalous scaling remains visible in the short-time dynamics.}
\label{fig:fig_msd_Pe}
\end{figure}
To disentangle the effects of viscoelastic memory from those of activity strength, we now examine the dependence of the mean-square displacement on the activity strength $\mathrm{Pe}$ for a non-circling swimmer ($\tilde{\omega}_0=0$).

Figure~\ref{fig:fig_msd_Pe} shows the mean-square displacement
$\langle\Delta\tilde{\mathbf r}^2(\tilde t)\rangle$ and the corresponding
dynamical exponent $\beta(\tilde t)$ for three activity strengths
$\mathrm{Pe}=1,40,100$ for a non-circling swimmer ($\tilde{\omega}_0=0$).
The governing MSD expression in Eq.~(\ref{eq:msd_total}) shows that at short times, the translational fractional memory term dominates, producing the anomalous scaling
$\langle\Delta\tilde{\mathbf r}^2\rangle \sim \tilde t^{\alpha_T}$ with
$\beta(\tilde t)\simeq\alpha_T$. Since the fractional term carries the prefactor $\mathrm{Pe}^{-2}$, increasing activity shifts the crossover time $\tilde t_{c1}$ to smaller
values. Consequently, the fractional regime becomes progressively compressed as $\mathrm{Pe}$ increases, as seen in Figs.~\ref{fig:fig_msd_Pe}(a,c,e) and in the corresponding dynamical
exponents in panels (b,d,f). For $\tilde t>\tilde t_{c1}$ the motion enters the ballistic regime $\langle\Delta\tilde{\mathbf r}^2\rangle\sim\tilde t^2$ with $\beta\simeq2$, arising from persistent active propulsion. As the activity increases, this ballistic window expands and dominates a larger portion of the intermediate-time dynamics. At long times, the orientational correlations decay and the motion crosses over to an effective diffusion, $\langle\Delta\tilde{\mathbf r}^2\rangle\sim 2D_{\rm act}\tilde t$, so that $\beta(\tilde t)\to1$. In contrast to the first crossover, the second crossover time $\tilde t_{c2}$ depends only on the rotational correlation function and is therefore essentially independent of $\mathrm{Pe}$. Thus increasing activity primarily suppresses the short-time fractional window while broadening the ballistic regime, whereas the anomalous exponent itself remains controlled solely by the viscoelastic memory parameter $\alpha$.

\subsection{Memory delay function and phase diagram of coupled memory}
\begin{figure}[t]
\centering
\includegraphics[width=\columnwidth]{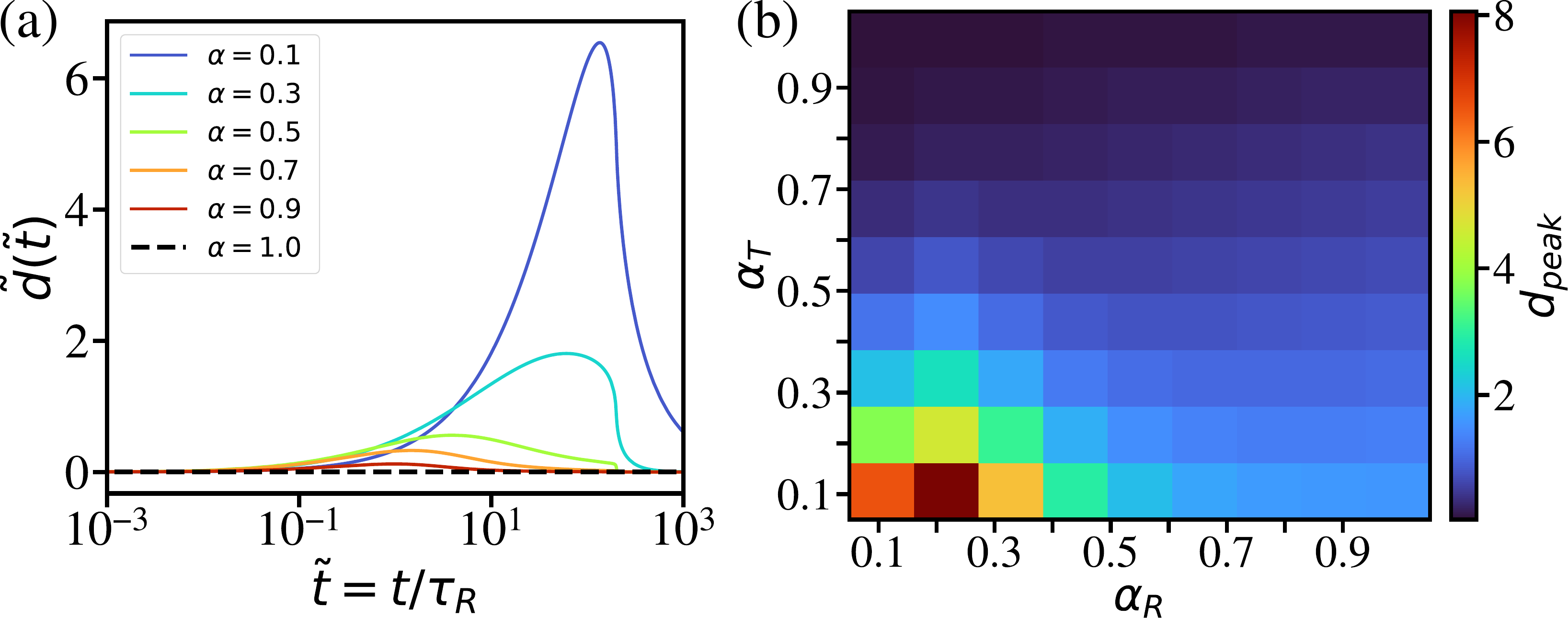}
\caption{(Color online) (a) Delay function $\tilde d(\tilde t)$ for different memory exponents $\alpha=\alpha_R=\alpha_T$. 
The delay vanishes in the Newtonian limit ($\alpha=1$) and develops a pronounced peak as memory strengthens. 
(b) Peak delay $d_{\mathrm{peak}}=\max[\tilde d(\tilde t)]$ in the $(\alpha_R,\alpha_T)$ plane. 
The delay is maximized when both translational and rotational memory are strong, demonstrating cooperative enhancement of force-orientation lag.}
\label{fig:memory_delay}
\end{figure}
We finally introduce a direct measure of non-Markovian coupling: the memory delay function $d(t)$, which quantifies the time-lag between the orientation and the effective propulsion force.
We discuss how this lag emerges from the combined action of translational and rotational memory. The memory delay function is defined as~\cite{Sprenger2022},
\begin{equation*}
d(t)
=
\langle \mathbf{F}_v(t) \cdot \hat{n}(0) \rangle
-
\langle \mathbf{F}_v(0) \cdot \hat{n}(t) \rangle,
\tag{9}
\end{equation*}
where the effective self-propulsion force is
\begin{equation*}
\mathbf{F}_v(t)
=
v_0 \int_{-\infty}^{t}
\Gamma_T(t-t')\,\hat{n}(t')\,dt'.
\quad
\end{equation*}
Physically, $d(t)$ measures the asymmetry between two projections:
the projection of the current effective force onto the initial orientation,
and the projection of the initial effective force onto the current orientation. In a Newtonian fluid, where $\Gamma_T(t)\propto \delta(t)$, the propulsion force is instantaneous and proportional to $\hat{n}(t)$. 
As a result, both projections are identical and $d(t)=0 \, \text{for all } t$. Thus, any nonzero delay function is a direct and unambiguous signature of viscoelastic memory. The delay function is evaluated in the stationary regime, in which the memory integral defining $\mathbf F_v(t)$ extends from $-\infty$ to $t$ and correlations depend only on the time difference. Accordingly, $C$ is continued to negative arguments through its even stationary form
\begin{equation}
C(\Delta t)=\cos(\tilde\omega_0\,\Delta t)\,\exp\!\left[-\frac{|\Delta t|^{\alpha_R}}{\Gamma(1+\alpha_R)}\right],
\tag{9a}
\label{eq:C_even}
\end{equation}
so that the term $C(\tilde t-\tilde\tau)$ appearing below is well defined for $\tilde\tau>\tilde t$. Using the derivation in Appendix~\ref{sec:memory_delay_appendix}, the dimensionless delay function reads
\begin{equation}
\tilde d(\tilde t)
=
\frac{1}{\Gamma(1-\alpha_T)}
\int_0^\infty
\tilde\tau^{-\alpha_T}
\big[
C(\tilde t-\tilde\tau)
-
C(\tilde t+\tilde\tau)
\big]
d\tilde\tau,
\tag{10}
\label{eq:delay}
\end{equation}
where the translational power-law kernel $\tilde\tau^{-\alpha_T}$ weights past orientations, while the rotational memory enters through the stretched-exponential correlation function
$C(\tilde t)$. 
A nonzero delay emerges only when both memory channels are present. 
Fig.~\ref{fig:memory_delay}(a) shows $\tilde d(\tilde t)$ for different memory exponents $\alpha=\alpha_R=\alpha_T$. 
In the Newtonian limit, the kernel is local and $\tilde d(\tilde t)=0$. 
For $\alpha<1$, a positive peak develops, indicating a finite time-lag between the propulsion force and the orientation of the swimmer. 
The peak grows as $\alpha$ decreases, reflecting the stronger contribution of past orientations to the instantaneous effective force. 
After reaching its maximum, the delay decays as orientational correlations relax, and the lag disappears once rotational coherence is lost. 
To quantify the strength of this effect, we define the peak delay $d_{\mathrm{peak}} = \max\,[\tilde d(\tilde t)]$, which characterizes the maximal force-orientation lag induced by memory. Figure~\ref{fig:memory_delay}(b) shows $d_{\mathrm{peak}}$ in the $(\alpha_R,\alpha_T)$ plane. The delay is generated by translational nonlocality: as $\alpha_T\to1$ the translational kernel becomes local, $\Gamma_T(\tau)\to c_T\delta(\tau)$, and Eq.~(\ref{eq:delay}) gives $\tilde d\propto\tfrac12[C(\tilde t)-C(\tilde t)]=0$ identically, so the delay vanishes exactly along the entire $\alpha_T=1$ edge. As $\alpha_R\to1$ the delay is instead strongly suppressed---the rotational correlation loses its heavy tail and the weighted history of past orientations shrinks---but it does not vanish identically for $\alpha_T<1$. The largest delays therefore occur when both exponents are small, demonstrating a cooperative enhancement of force-orientation delay when strong translational memory acts together with long-lived rotational correlations. Thus $d_{\mathrm{peak}}$
provides a quantitative measure of the strength of non-Markovian coupling between propulsion and orientation.

\section{Conclusion}
\label{sec:conclusion}
In this work, we have developed a framework for the dynamics of an active particle in a viscoelastic medium with long-range power-law memory~\cite{Metzler2000, Goychuk2012}. By employing a Generalized Langevin Equation with a fractional memory kernel, our model moves beyond the single-timescale relaxation of the canonical Maxwell models~\cite{Sprenger2022, Sevilla2019} and the memoryless limit of the standard Active Brownian Particle~\cite{Fily2012}, providing a more realistic description of complex soft matter systems~\cite{Metzler2000, Barkai2012, Metzler2014, Granick1995soft, Guigas2007, Weber2012nonthermal, Solon2015}. 

We analytically solved the model and found that the orientational correlation function exhibits stretched-exponential decay, implying that orientational persistence is governed by a broad spectrum of relaxation times rather than a single characteristic timescale. 
Translational memory modifies the short-time transport. 
Instead of the diffusive onset $\mathrm{MSD}\sim t$ characteristic of Newtonian active Brownian particles~\cite{Romanczuk2012,Fily2012}, the motion displays fractional scaling $\mathrm{MSD}\sim t^{\alpha_T}$, providing a clear dynamical signature of viscoelastic memory. 
At intermediate times, persistent self-propulsion produces ballistic motion, while the long-time dynamics becomes effectively diffusive once orientational correlations decay. 
Our analysis shows that viscoelastic memory controls the anomalous scaling exponents and persistence of orientational correlations, whereas the translational activity primarily determines the extent of the ballistic regime and the crossover structure between the dynamical regimes. 
For chiral swimmers, the addition of memory enhances the oscillatory regime with coherent spiral motion. 
A distinctive hallmark of the non-Markovian dynamics is the emergence of a time delay in the force-orientation alignment. 
The power-law memory kernel generates a finite delay because the effective propulsion force depends on a weighted history of past orientations. 
The peak delay $d_{\mathrm{peak}}$ provides a quantitative measure of non-Markovian coupling strength and reveals a cooperative enhancement region in the $(\alpha_R,\alpha_T)$ plane. 

Our results establish that viscoelastic memory qualitatively reshapes active transport by introducing fractional short-time dynamics, memory-controlled crossover behavior, enhanced persistence of chiral motion, and lag between propulsion and orientation. 
Connecting a physically motivated memory kernel to observable dynamics provides a tool for interpreting experimental data from particle tracking~\cite{Waigh2005, Squires2010} and for inferring the rheological properties of a medium from such measurements. 
The approach is applicable to describe microswimmers in complex fluids~\cite{Patteson2016active, Saintillan2018} such as bacteria navigating mucus~\cite{Patteson2015}, immune cells migrating through the extracellular matrix~\cite{Hallmann2015regulation}, or synthetic active colloids in dense polymer networks~\cite{Zottl2019, Narinder2018}. 

Two simplifications of the present model deserve comment. First, we treat the self-propulsion speed as constant. In real microswimmers the propulsion force fluctuates and may even oscillate on time scales comparable to the bath relaxation---for instance, propulsion-force correlations decay over seconds in Salmonella and algae exhibit force oscillations with periods of $\sim20$~ms~\cite{Klimek2026}. Incorporating a fluctuating speed would couple the speed and orientation correlations in the active mean-square displacement and is a natural extension of the present exactly solvable model. Second, we work in two dimensions for analytical transparency; besides being a tractable idealization, this directly describes swimmers confined to a plane---near interfaces and substrates, in thin films, and in the confined pore geometries of soil inhabited by certain algae~\cite{Klimek2026}. The method generalizes to three dimensions, where rotational relaxation on the sphere modifies the explicit form of $C(\tilde t)$ while leaving the qualitative phenomenology---memory-controlled persistence, fractional short-time transport, and the force--orientation delay---intact.
Our model also offers a theoretical benchmark for experiments and simulations on collective behavior, particle-boundary interactions, and the response to external fields in media with power-law viscoelasticity~\cite{Crocker2000, Loverdo2008, Jeon2011}, and stimulates future research on novel forms of association~\cite{Gomez-Solano2016, Cates2015, Bialke2013}, including in spatial confinement~\cite{Elgeti2015, Ahana2019confinement, Das2018}, with hydrodynamic interactions~\cite{Lauga2007,ouyang2023swimming, Spagnolie2012}, or materials with yielding or shear-thinning properties~\cite{Puertas2014, Mckinley2018anomalous}.

\section{Acknowledgments}
M.K. acknowledges funding support from the University of Chinese Academy of Sciences (UCAS) and Wenzhou Institute of the University of Chinese Academy of Sciences (WIUCAS). M.K. also acknowledges the computational facilities provided by WIUCAS and the Departament de F\'{\i}sica de la Mat\`{e}ria Condensada, Universitat de Barcelona.
I.P. acknowledges DURSI for financial support under Project No. 2021SGR-673, Ministerio de Ciencia, Innovaci\'on y Universidades MCIU/AEI/FEDER for financial support under grant agreement PID2024-156516NB-100 AEI/FEDER-EU, and Generalitat de Catalunya for financial support under Program Icrea Acad\`emia. 
J.D. acknowledges the support of Slovenian Research Agency through the strategic project 21-ARRS STR-0002.

\section{Data availability}
The scripts used to evaluate the analytical expressions, perform the numerical quadratures, and generate all figures are openly available at the following \href{https://doi.org/10.5281/zenodo.21441715}{\texttt{[URL/DOI: https://doi.org/10.5281/zenodo.21441715]}}.

\begin{widetext}
\appendix
\section{Detailed derivations of statistical observables}
\label{sec:appendix_derivations}

This appendix provides a detailed derivation of the main statistical quantities discussed in the paper. We start from the generalized Langevin equations (GLEs) for a particle with power-law memory and use the Laplace transform method to find the solutions for the key observables in their dimensional form. Subsequently, each solution is non-dimensionalized to arrive at the final form used in the main text.

\subsection{Model formulation and preliminaries}
\label{subsec:model_appendix}

The dynamics of the particle are governed by a set of coupled, non-Markovian GLEs for the translational and rotational degrees of freedom.
\begin{align}
\int_{0}^{t} \Gamma_T(t-t') \left( \dot{\mathbf{r}}(t') - v_0 \hat{n}(t') \right) dt' &= \boldsymbol{\xi}(t) \label{eq:translational_gle_appendix} \\
\int_{0}^{t} \Gamma_R(t-t') \left( \dot{\phi}(t') - \omega_0 \right) dt' &= \eta(t) \label{eq:rotational_gle_appendix}
\end{align}
The central feature of our model is the power-law memory kernel:
\begin{equation}
\Gamma_{k}(t) = \frac{c_{k}}{\Gamma(1-\alpha_{k})} t^{-\alpha_{k}} \quad \text{for } t > 0, \text{ and } 0 < \alpha_k < 1
\label{eq:memory_kernel_appendix}
\end{equation}
where the subscript $k$ stands for either $T$ (translational) or $R$ (rotational) and $c_k\equiv c_{\alpha_k}$ is the fractional friction strength. It carries exponent-dependent units. The stochastic forces are zero-mean Gaussian processes whose correlations are given by the Fluctuation-Dissipation Theorem (FDT):
\begin{align}
\langle \xi_i(t) \xi_j(t') \rangle &= k_B T \Gamma_T(|t-t'|) \delta_{ij} \label{eq:fdt_trans_appendix} \\
\langle \eta(t) \eta(t') \rangle &= k_B T \Gamma_R(|t-t'|) \label{eq:fdt_rot_appendix}
\end{align}
The Laplace transform of the power-law kernel, a crucial tool for our derivations, is:
\begin{equation}
\tilde{\Gamma}_k(s) = \mathcal{L}\left\{\frac{c_k}{\Gamma(1-\alpha_k)} t^{-\alpha_k}\right\} = \frac{c_k}{\Gamma(1-\alpha_k)} \frac{\Gamma(1-\alpha_k)}{s^{1-\alpha_k}} = c_k s^{\alpha_k-1}
\label{eq:laplace_kernel_appendix}
\end{equation}

\subsection{Solution for the particle orientation $\phi(t)$}
First, we solve the rotational GLE (Eq. \ref{eq:rotational_gle_appendix}) for the orientation angle $\phi(t)$. Applying the Laplace transform and using the convolution theorem gives:
\begin{equation}
\tilde{\Gamma}_R(s) \mathcal{L}\{\dot{\phi}(t) - \omega_0\} = \tilde{\eta}(s)
\end{equation}
Using $\mathcal{L}\{\dot{\phi}(t)\} = s \tilde{\phi}(s) - \phi(0)$ and $\mathcal{L}\{\omega_0\} = \omega_0 / s$, and substituting the form of $\tilde{\Gamma}_R(s)$ from Eq. \ref{eq:laplace_kernel_appendix}:
\begin{equation}
c_R s^{\alpha_R-1} \left( s\tilde{\phi}(s) - \phi_0 - \frac{\omega_0}{s} \right) = \tilde{\eta}(s)
\end{equation}
We rearrange to solve for $\tilde{\phi}(s)$:
\begin{align}
s^{\alpha_R} \tilde{\phi}(s) - s^{\alpha_R-1}\phi_0 - s^{\alpha_R-1}\frac{\omega_0}{s} &= \frac{\tilde{\eta}(s)}{c_R} \nonumber \\
\tilde{\phi}(s) &= \frac{\phi_0}{s} + \frac{\omega_0}{s^2} + \frac{s^{-\alpha_R}}{c_R} \tilde{\eta}(s)
\end{align}
Taking the inverse Laplace transform term-by-term, we recognize that $s^{-\alpha_R}$ corresponds to convolution with the function $t^{\alpha_R-1}/\Gamma(\alpha_R)$, which represents a fractional integral. The solution for $\phi(t)$ is therefore:
\begin{equation}
\phi(t) = \phi_0 + \omega_0 t + \frac{1}{c_R} \int_0^t \frac{(t-t')^{\alpha_R-1}}{\Gamma(\alpha_R)} \eta(t') dt'
\label{eq:phi_time_appendix}
\end{equation}
This shows that the orientation is the initial angle, plus a deterministic drift, plus a stochastic term arising from the fractionally integrated noise.

\subsection{Derivation of the orientational correlation function $C(t)$}
\label{sec:correlation_derivation_appendix}

The orientational correlation function is defined as $C(t) = \langle \hat{n}(t) \cdot \hat{n}(0) \rangle = \langle \cos(\phi(t) - \phi_0) \rangle$. Using the solution for $\phi(t)$ from Eq. \ref{eq:phi_time_appendix}, the angular displacement is $\Delta\phi(t) = \omega_0 t + \Delta\phi_{stoch}(t)$, where $\Delta\phi_{stoch}(t)$ is the stochastic integral term. Since $\eta(t)$ is a zero-mean Gaussian process, so is $\Delta\phi_{stoch}(t)$.

Using the identity $\langle e^{iX} \rangle = e^{-\frac{1}{2} \langle X^2 \rangle}$ for a zero-mean Gaussian variable $X$:
\begin{align}
C(t) &= \langle \cos(\omega_0 t + \Delta\phi_{stoch}(t)) \rangle = \text{Re} \left[ e^{i\omega_0 t} \langle e^{i\Delta\phi_{stoch}(t)} \rangle \right] \nonumber \\
&= \cos(\omega_0 t) \exp\left( -\frac{1}{2}\langle (\Delta\phi_{stoch}(t))^2 \rangle \right) = \cos(\omega_0 t) \exp\left( -\frac{1}{2} M_2(t) \right)
\end{align}
where $M_2(t) = \langle (\Delta\phi_{stoch}(t))^2 \rangle$ is the mean square angular displacement. From the FDT, its Laplace transform is $\tilde{M}_2(s) = \frac{2 k_B T}{s^2 \tilde{\Gamma}_R(s)}$. Substituting Eq. \ref{eq:laplace_kernel_appendix}:
\begin{equation}
\tilde{M}_2(s) = \frac{2 k_B T}{s^2 (c_R s^{\alpha_R-1})} = \frac{2 k_B T}{c_R} s^{-1-\alpha_R}
\end{equation}
Performing the inverse Laplace transform using $\mathcal{L}^{-1} \{ s^{-\nu} \} = t^{\nu-1} / \Gamma(\nu)$:
\begin{equation}
M_2(t) = \frac{2 k_B T}{c_R \Gamma(1+\alpha_R)} t^{\alpha_R}
\end{equation}
Substituting this back into the expression for $C(t)$, we obtain the final dimensional form:
\begin{equation}
C(t) = \cos(\omega_0 t) \exp\left( - \frac{k_B T}{c_R \Gamma(1+\alpha_R)} t^{\alpha_R} \right)
\label{eq:corr_final_appendix}
\end{equation}

\subsubsection{Dimensionless form of $C(t)$}
We introduce the rotational fractional time scale $\tau_R \equiv \left(\frac{c_R}{k_BT}\right)^{1/\alpha_R}$ and dimensionless frequency $\tilde{\omega}_0 = \omega_0 \tau_R$. Substituting $t = \tau_R \tilde{t}$ into Eq. \ref{eq:corr_final_appendix}:
\begin{align}
C(\tilde{t}) &= \cos\left(\frac{\tilde{\omega}_0}{\tau_R} \tau_R \tilde{t}\right) \exp\left( - \frac{k_B T}{c_R \Gamma(1+\alpha_R)} (\tau_R \tilde{t})^{\alpha_R} \right) \nonumber \\
&= \cos(\tilde{\omega}_0 \tilde{t}) \exp\left( - \frac{1}{\tau_R^{\alpha_R} \Gamma(1+\alpha_R)} \tau_R^{\alpha_R} \tilde{t}^{\alpha_R} \right)
\end{align}
This gives the final dimensionless equation refered in the main text:
\begin{equation}
\boxed{C(\tilde{t}) = \cos(\tilde{\omega}_0 \tilde{t}) \exp\left( - \frac{\tilde{t}^{\alpha_R}}{\Gamma(1+\alpha_R)}  \right)}
\label{eq:C_dimensionless_final}
\end{equation}

\subsection{Derivation of the mean displacement $\langle \Delta \mathbf{r}(t) \rangle$}
\label{sec:mean_displacement_appendix}

Taking the ensemble average of the translational GLE (Eq. \ref{eq:translational_gle_appendix}) makes the noise term vanish: $\langle \boldsymbol{\xi}(t) \rangle = 0$.
\begin{equation}
\int_{0}^{t} \Gamma_T(t-t') \left( \langle \dot{\mathbf{r}}(t') \rangle - v_0 \langle \hat{n}(t') \rangle \right) dt' = \mathbf{0}
\end{equation}
Applying the Laplace transform and using the convolution theorem, we get:
\begin{equation}
\tilde{\Gamma}_T(s) \left( \mathcal{L}\{\langle \dot{\mathbf{r}}(t) \rangle\}(s) - v_0 \mathcal{L}\{\langle \hat{n}(t) \rangle\}(s) \right) = \mathbf{0}
\end{equation}
Since $\tilde{\Gamma}_T(s) \neq 0$, this implies $\mathcal{L}\{\langle \dot{\mathbf{r}}(t) \rangle\}(s) = v_0 \mathcal{L}\{\langle \hat{n}(t) \rangle\}(s)$. Using $\mathcal{L}\{\langle \dot{\mathbf{r}}(t) \rangle\} = s \langle \tilde{\Delta \mathbf{r}}(s) \rangle$, we find $\langle \tilde{\Delta \mathbf{r}}(s) \rangle = \frac{v_0}{s} \langle \tilde{\hat{n}}(s) \rangle$. The factor of $1/s$ corresponds to time integration, so the dimensional solution is:
\begin{equation}
\langle \Delta \mathbf{r}(t) \rangle = v_0 \int_0^t \langle \hat{n}(t') \rangle dt'
\end{equation}
The average orientation vector is $\langle \hat{n}(t) \rangle = \langle (\cos \phi(t), \sin \phi(t)) \rangle$. Writing $\phi(t)=\phi_0+\omega_0 t+\Delta\phi_{stoch}(t)$ and using the Gaussian identity $\langle e^{i\Delta\phi_{stoch}}\rangle=e^{-\frac12 M_2(t)}$ exactly as for $C(t)$, we obtain
\begin{equation}
\langle \hat{n}(t) \rangle = E(t)\,\big(\cos(\phi_0 + \omega_0 t), \sin(\phi_0 + \omega_0 t)\big),
\qquad
E(t)\equiv e^{-\frac12 M_2(t)}=\exp\!\Big[-\tfrac{k_BT}{c_R\Gamma(1+\alpha_R)}t^{\alpha_R}\Big].
\end{equation}
We stress that the relevant decay factor is the envelope $E(t)$, which is distinct from the scalar two-time correlation $C(t)=\cos(\omega_0 t)E(t)$: the mean orientation rotates as $\cos(\phi_0+\omega_0 t)$ and decays as $E(t)$, so the extra oscillatory factor $\cos(\omega_0 t)$ carried by $C(t)$ must not be reused here.

\subsubsection{Dimensionless mean displacement}
We non-dimensionalize the displacement by the persistence length $L_p = v_0 \tau_R$.
\begin{align}
\langle \Delta \tilde{\mathbf{r}}(\tilde{t}) \rangle &= \frac{\langle \Delta \mathbf{r}(t) \rangle}{L_p} = \frac{1}{v_0 \tau_R} \left( v_0 \int_0^{\tau_R \tilde{t}} \langle \hat{n}(t') \rangle dt' \right) \nonumber \\
&= \frac{1}{\tau_R} \int_0^{\tilde{t}} \langle \hat{n}(\tau_R \tilde{t}') \rangle (\tau_R d\tilde{t}') = \int_0^{\tilde{t}} \langle \hat{n}(\tilde{t}') \rangle d\tilde{t}'
\end{align}
where $\langle \hat{n}(\tilde{t}') \rangle$ uses the dimensionless envelope $E(\tilde t')=\exp[-\tilde t'^{\alpha_R}/\Gamma(1+\alpha_R)]$. This gives the final dimensionless equation:
\begin{equation}
\boxed{\langle \Delta \tilde{\mathbf{r}}(\tilde{t}) \rangle = \int_0^{\tilde{t}} E(\tilde{t}') \left( \cos(\phi_0 + \tilde{\omega}_0 \tilde{t}') \hat{i} + \sin(\phi_0 + \tilde{\omega}_0 \tilde{t}') \hat{j} \right) d\tilde{t}'}
\end{equation}

\subsection{Derivation of the mean-square displacement $\text{MSD}(t)$}
\label{sec:msd_appendix}

The MSD is the sum of a thermal and an active part, $\text{MSD}(t) = \text{MSD}_{th}(t) + \text{MSD}_a(t)$.

\subsubsection{Thermal part}
The derivation is identical to the rotational case, but for 2D translational motion, giving $\tilde{\text{MSD}}_{th}(s) = \frac{4 k_B T}{s^2 \tilde{\Gamma}_T(s)}$. Inverting the Laplace transform gives:
\begin{equation}
\text{MSD}_{th}(t) = \frac{4 k_B T}{c_T \Gamma(1+\alpha_T)} t^{\alpha_T}
\end{equation}

\subsubsection{Active part}
This is the double time integral of the active velocity autocorrelation function, $C_{v,a}(t) = \langle v_0 \hat{n}(t) \cdot v_0 \hat{n}(0) \rangle = v_0^2 C(t)$.
\begin{equation}
\text{MSD}_a(t) = 2 \int_0^t (t-\tau) v_0^2 C(\tau) d\tau
\end{equation}
The total dimensional MSD is:
\begin{equation}
\text{MSD}(t) = \frac{4 k_B T t^{\alpha_T}}{c_T \Gamma(1+\alpha_T)} + 2 v_0^2 \int_0^t (t-\tau) C(\tau) d\tau
\label{eq:msd_total_final_appendix}
\end{equation}

\subsubsection{Dimensionless MSD}
We non-dimensionalize by $L_p^2 = (v_0 \tau_R)^2$.

\paragraph{Thermal Part:}
\begin{align}
\text{MSD}_{th}(\tilde t) &= \frac{4 k_B T (\tau_R \tilde{t})^{\alpha_T}}{c_T \Gamma(1+\alpha_T) (v_0 \tau_R)^2} = \frac{4k_BT\,\tau_R^{\alpha_T-2}}{c_T\,\Gamma(1+\alpha_T)\,v_0^2}\,\tilde t^{\alpha_T}
\end{align}
This suggests the dimensionless P\'eclet number
$\mathrm{Pe}^2 \equiv \frac{v_0^2\,c_T\,\tau_R^{2-\alpha_T}}{k_BT}$, this becomes:
\begin{equation}
\text{MSD}_{th}(\tilde t) = \frac{4}{\text{Pe}^2\Gamma(1+\alpha_T)} \tilde{t}^{\alpha_T}
\end{equation}

\paragraph{Active Part:}
\begin{equation}
\text{MSD}_{a}(\tilde t) = \frac{2 v_0^2}{(v_0 \tau_R)^2} \int_0^{\tau_R \tilde{t}} \tau_R (\tilde{t}-\tilde{\tau}) C(\tau) d\tau = 2 \int_0^{\tilde{t}} (\tilde{t}-\tilde{\tau}) C(\tilde{\tau}) d\tilde{\tau}
\end{equation}
The total dimensionless MSD is:
\begin{equation}
\boxed{\langle\Delta\tilde{\mathbf{r^2}}(\tilde t)\rangle = \frac{4}{\text{Pe}^2 \Gamma(1+\alpha_T)} \tilde{t}^{\alpha_T} + 2 \int_0^{\tilde{t}} (\tilde{t}-\tilde{\tau}) C(\tilde{\tau}) d\tilde{\tau}}\label{dimless_MSD}
\end{equation}

\subsection{Derivation of the memory delay function $d(t)$}
\label{sec:memory_delay_appendix}

\subsubsection{From definition to integral form}
The memory delay function is defined as $d(t) = \langle \mathbf{F}_v(t) \cdot \hat{n}(0) \rangle - \langle \mathbf{F}_v(0) \cdot \hat{n}(t) \rangle$, where the effective self-propulsion force is $\mathbf{F}_v(t) = v_0 \int_{-\infty}^{t} \Gamma_T(t-t') \hat{n}(t') dt'$.

The definition of $\mathbf{F}_v(t)$ assumes the system is in a stationary state, where correlations depend only on time differences. The orientational correlation function $C(t)$ derived for the system starting at $t=0$ has the same functional form as the stationary-state correlation function, allowing us to use it in this context.

We evaluate the first term, using the stationary property $\langle \hat{n}(t') \cdot \hat{n}(0) \rangle = C(t')$.
\begin{align}
\langle \mathbf{F}_v(t) \cdot \hat{n}(0) \rangle &= \left\langle \left( v_0 \int_{-\infty}^{t} \Gamma_T(t-t') \hat{n}(t') dt' \right) \cdot \hat{n}(0) \right\rangle \nonumber \\
&= v_0 \int_{-\infty}^{t} \Gamma_T(t-t') \langle \hat{n}(t') \cdot \hat{n}(0) \rangle dt' \nonumber \\
&= v_0 \int_{-\infty}^{t} \Gamma_T(t-t') C(t') dt'
\end{align}
Now, we perform a change of variables, letting $\tau = t - t'$, which means $t' = t - \tau$ and $dt' = -d\tau$. The integration limits change from $(-\infty, t)$ for $t'$ to $(\infty, 0)$ for $\tau$.
\begin{equation}
= v_0 \int_{\infty}^{0} \Gamma_T(\tau) C(t-\tau) (-d\tau) = v_0 \int_{0}^{\infty} \Gamma_T(\tau) C(t-\tau) d\tau
\end{equation}
Now, we evaluate the second term, using $\langle \hat{n}(t') \cdot \hat{n}(t) \rangle = C(t-t')$.
\begin{align}
\langle \mathbf{F}_v(0) \cdot \hat{n}(t) \rangle &= \left\langle \left( v_0 \int_{-\infty}^{0} \Gamma_T(0-t') \hat{n}(t') dt' \right) \cdot \hat{n}(t) \right\rangle \nonumber \\
&= v_0 \int_{-\infty}^{0} \Gamma_T(-t') \langle \hat{n}(t') \cdot \hat{n}(t) \rangle dt' \nonumber \\
&= v_0 \int_{-\infty}^{0} \Gamma_T(-t') C(t-t') dt'
\end{align}
We perform another change of variables, letting $\tau = -t'$, which means $t' = -\tau$ and $dt' = -d\tau$. The limits change from $(-\infty, 0)$ for $t'$ to $(\infty, 0)$ for $\tau$.
\begin{equation}
= v_0 \int_{\infty}^{0} \Gamma_T(\tau) C(t-(-\tau)) (-d\tau) = v_0 \int_{0}^{\infty} \Gamma_T(\tau) C(t+\tau) d\tau
\end{equation}
Combining the two results, we arrive at the integral form:
\begin{equation}
d(t) = v_0 \int_{0}^{\infty} \Gamma_T(\tau) \left[ C(t-\tau) - C(t+\tau) \right] d\tau
\label{eq:delay_integral_appendix}
\end{equation}

\subsubsection{Dimensionless delay function}
Insert $\Gamma_T(\tau)=c_T \tau^{-\alpha_T}/\Gamma(1-\alpha_T)$ and rescale
$\tau=\tau_R\tilde\tau$, $t=\tau_R\tilde t$. This yields
\begin{equation}
d(t)=\frac{v_0\,c_T\,\tau_R^{1-\alpha_T}}{\Gamma(1-\alpha_T)}
\int_0^\infty \tilde\tau^{-\alpha_T}\Big[C(\tilde t-\tilde\tau)-C(\tilde t+\tilde\tau)\Big]\,d\tilde\tau.
\end{equation}
Therefore, dimensionless memory delay function $\tilde d(\tilde t)$ defined as:
\begin{equation}
\boxed{
\tilde d(\tilde t)\equiv \frac{d(t)}{v_0 c_T \tau_R^{1-\alpha_T}}
=\frac{1}{\Gamma(1-\alpha_T)}
\int_0^\infty \tilde\tau^{-\alpha_T}\Big[C(\tilde t-\tilde\tau)-C(\tilde t+\tilde\tau)\Big]\,d\tilde\tau,
}
\label{eq:delay_dimensionless_final}
\end{equation}
with $C(\tilde t)$ given by \eqref{eq:C_dimensionless_final}.

\section{Orientation correlation \& mean displacement for non-circling swimmer}
\label{sec:appendix_noncircling}
\begin{figure}[!ht]
\centering
\includegraphics[width=0.8\columnwidth]{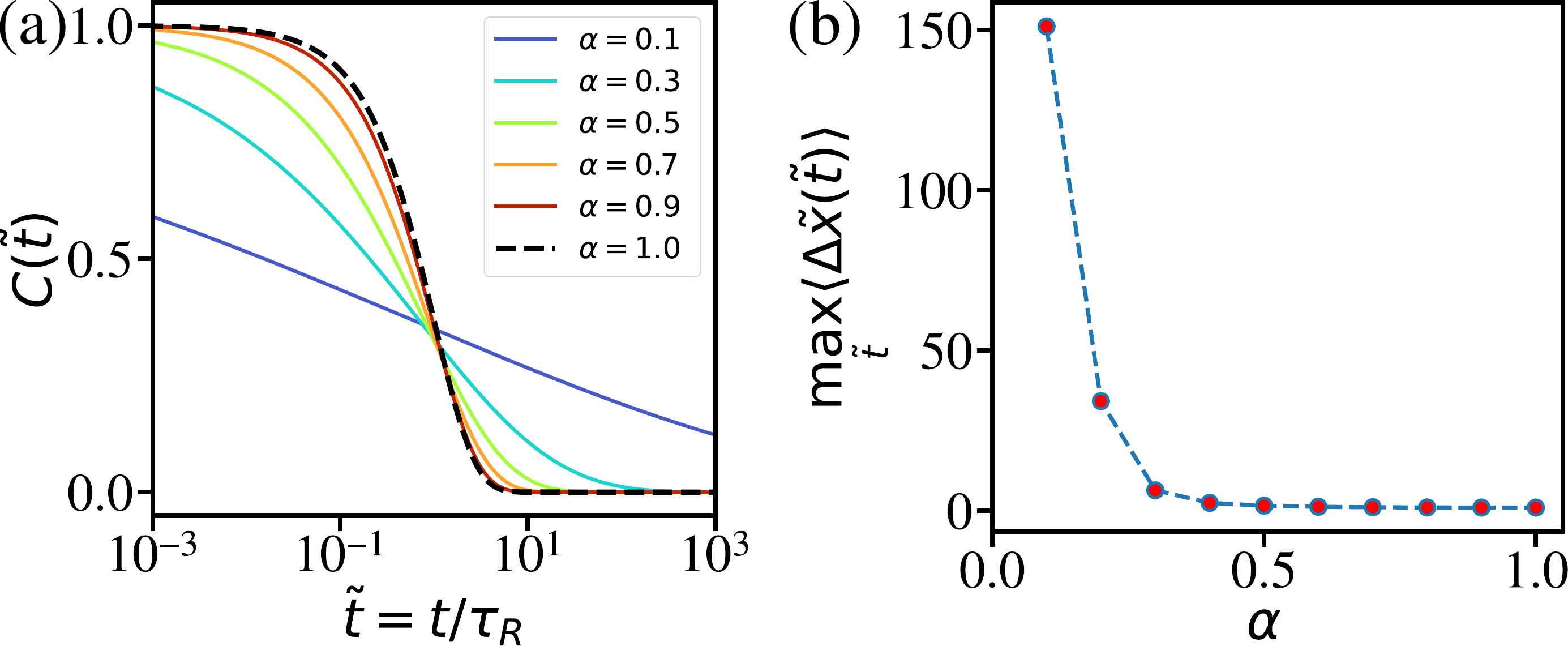}
\caption{(Color online) Non-circling swimmer ($\tilde{\omega}_0=0$). 
(a) Orientational correlation function $C(\tilde t)$ for different memory exponents $\alpha=\alpha_R=\alpha_T$. 
The Newtonian case ($\alpha=1$, dashed black line) exhibits exponential decay, while $\alpha<1$ produces stretched-exponential relaxation and long-lived persistence. 
(b) Maximum longitudinal mean displacement, $\max_{\tilde t}\langle \Delta \tilde x(\tilde t)\rangle$, as a function of $\alpha$. The strong increase at small $\alpha$ shows that enhanced orientational memory amplifies directed displacement along the initial orientation.}
\label{fig:fig_appendix}
\end{figure}

In this section we provide the explicit forms of the orientational correlation and mean displacement for a non-circling swimmer. For $\tilde{\omega}_0 = 0$, Eq.~\ref{eq:orientation} in the main text simplifies to
\begin{equation}
C(\tilde t)
=
\exp\!\left[
-\frac{\tilde t^{\alpha_R}}{\Gamma(1+\alpha_R)}
\right].
\label{eq:C_noncircling}
\end{equation}

Thus, for a non-circling swimmer, viscoelastic memory modifies only the decay envelope of persistence. 
In the Newtonian limit ($\alpha_R=1$), the correlation reduces to the familiar exponential form $C(\tilde t)=e^{-\tilde t}$. For $0<\alpha_R<1$, the decay becomes slower, leading to long-lived orientational coherence. Figure~\ref{fig:fig_appendix}(a) shows $C(\tilde t)$ for different $\alpha$. 
As $\alpha$ decreases, the decay becomes progressively slower, producing a heavy tail that reflects long-time rotational memory. The dimensionless mean displacement is given by
\begin{equation}
\langle \Delta\tilde{\mathbf r}(\tilde t) \rangle
=
\int_0^{\tilde t}
C(\tilde t')
\left(
\cos\phi_0\,\hat i
+
\sin\phi_0\,\hat j
\right)
d\tilde t'.
\label{eq:mean_disp_noncircling}
\end{equation}

Since $\phi(t)$ does not rotate on average when $\tilde{\omega}_0=0$, the mean orientation remains aligned with the initial direction. 
Therefore, the motion is strictly one-dimensional along $\hat n(0)$:
\begin{equation}
\langle \Delta\tilde{\mathbf r}(\tilde t) \rangle
=
\hat n(0)
\int_0^{\tilde t}
C(\tilde t')
\, d\tilde t'.
\end{equation}

Choosing $\phi_0=0$ for simplicity,
\begin{equation}
\boxed{
\langle \Delta\tilde x(\tilde t) \rangle
=
\int_0^{\tilde t}
\exp\!\left[
-\frac{\tilde t'^{\alpha_R}}{\Gamma(1+\alpha_R)}
\right]
d\tilde t',
\qquad
\langle \Delta\tilde y(\tilde t) \rangle = 0.
}
\label{eq:mean_disp_x_noncircling}
\end{equation}

Thus, for a non-circling swimmer, the trajectory remains strictly along the initial direction. At long times, the displacement saturates to a finite value,
\[
\langle \Delta\tilde x(\infty) \rangle
=
\int_0^{\infty}
\exp\!\left[
-\frac{\tilde t^{\alpha_R}}{\Gamma(1+\alpha_R)}
\right]
d\tilde t,
\]
which increases as $\alpha_R$ decreases due to enhanced orientational persistence.
Since the transverse mean displacement remains zero for all memory exponents, we do not show $\langle \Delta \tilde y\rangle$ in Fig.~\ref{fig:fig_appendix}(b). Instead, we quantify the memory dependence of the directed response by plotting the maximum longitudinal mean displacement, $\max_{\tilde t}\langle \Delta \tilde x(\tilde t)\rangle$, as a function of $\alpha$.

\section{Mean-square displacement for distinct memory exponents}
\label{sec:appendix_alphaTR}
The two memory exponents enter the dynamics through different channels: the translational exponent $\alpha_T$ controls the fractional short-time term of the MSD [first term of Eq.~(\ref{eq:msd_total})], while the rotational exponent $\alpha_R$ controls the orientational correlation $C(\tilde t)$ and hence the persistence, the ballistic window, and the long-time crossover. Figure~\ref{fig:appendix_alphaTR} illustrates this by plotting the MSD and the dynamical exponent $\beta(\tilde t)$ for several unequal pairs $(\alpha_T,\alpha_R)$ for a non-circling swimmer ($\tilde\omega_0=0$). As anticipated, the short-time exponent obeys $\beta(\tilde t\!\to\!0)\to\alpha_T$ irrespective of $\alpha_R$, whereas the intermediate ballistic plateau and its terminal crossover to diffusion are set by $\alpha_R$.

\begin{figure}[!ht]
\centering
\includegraphics[width=0.8\columnwidth]{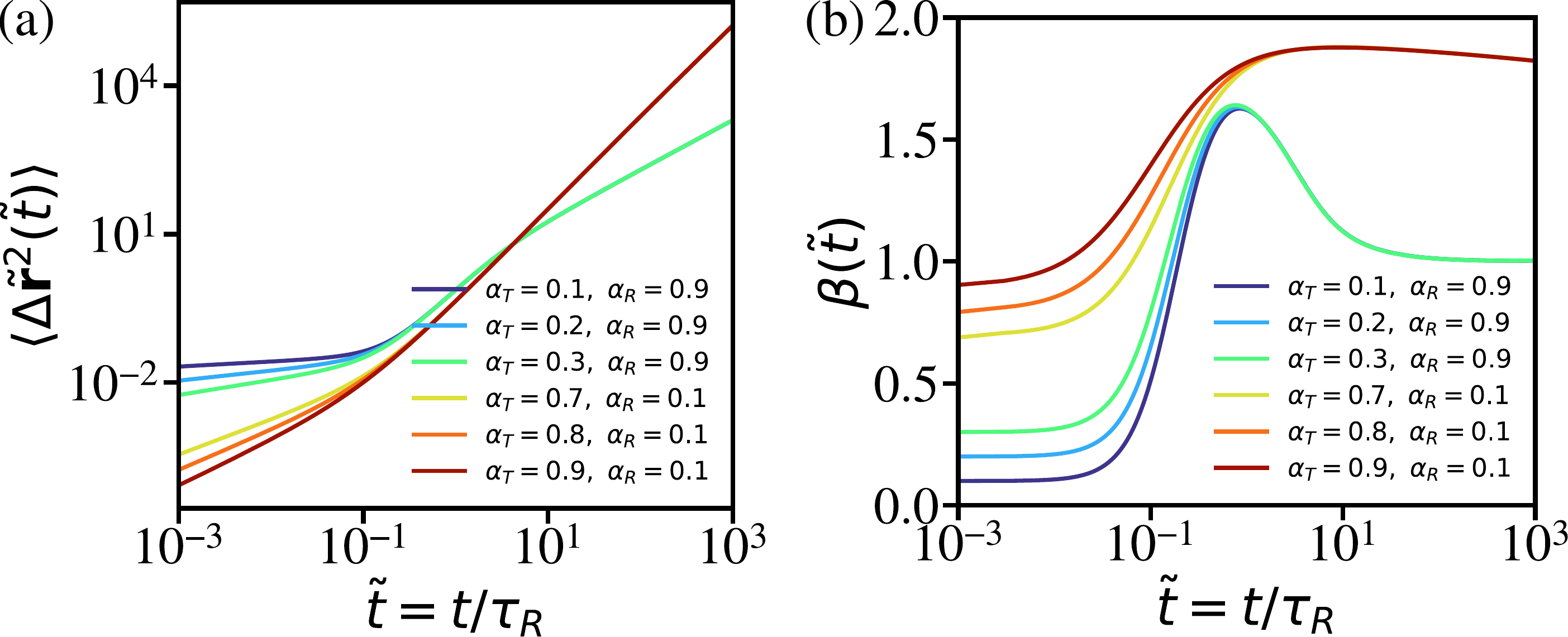}
\caption{(Color online) Mean-square displacement $\langle\Delta\tilde{\mathbf r}^2(\tilde t)\rangle$ (a) and dynamical exponent $\beta(\tilde t)$ (b) for a non-circling swimmer ($\tilde\omega_0=0$) with distinct translational and rotational memory exponents $(\alpha_T,\alpha_R)$, at $\mathrm{Pe}=10$. The short-time subdiffusive slope is set by $\alpha_T$ [$\beta(\tilde t\!\to\!0)\to\alpha_T$], while the extent of the ballistic regime and the long-time crossover to diffusion are governed by $\alpha_R$.}
\label{fig:appendix_alphaTR}
\end{figure}

\section{Asymptotic behavior of the MSD}\label{asymp_MSD}
To obtain the asymptotic behavior of the MSD, we first expand the orientational correlation at short times. We have
\begin{equation}
C(\tilde t)
\approx
1-a\tilde t^{\alpha_R}
-\frac{\tilde\omega_0^2}{2}\tilde t^2
+\frac{a^2}{2}\tilde t^{2\alpha_R}
+\cdots, 
\end{equation}

where $a=\frac{1}{\Gamma(1+\alpha_R)}$. Substituting this into the MSD integral in Eq.~(\ref{dimless_MSD}) and using
\[
\int_0^{\tilde t}(\tilde t-\tilde \tau)\tilde \tau^\beta\,d\tilde\tau
=
\frac{\tilde t^{\beta+2}}{(\beta+1)(\beta+2)},
\]
we obtain
\begin{equation}
\langle\Delta\tilde{\mathbf r}^2(\tilde t)\rangle
\approx
\frac{4\,\tilde t^{\alpha_T}}{\mathrm{Pe}^2\Gamma(1+\alpha_T)}
+\tilde t^2
-\frac{2\,\tilde t^{\alpha_R+2}}
{\Gamma(1+\alpha_R)(\alpha_R+1)(\alpha_R+2)}
+\cdots.
\end{equation}

For long times, since $C(\tilde t)=\cos(\tilde\omega_0\tilde t)e^{-a\tilde t^{\alpha_R}}$ is integrable, we write
\[
2\int_0^{\tilde t}(\tilde t-\tilde\tau)C(\tilde\tau)\,d\tilde\tau
=
2\tilde t\int_0^{\tilde t}C(\tilde\tau)\,d\tilde\tau
-2\int_0^{\tilde t}\tilde\tau C(\tilde\tau)\,d\tilde\tau.
\]
Defining
\[
D_{\rm act}=\int_0^\infty C(\tilde\tau)\,d\tilde\tau,
\qquad
B=\int_0^\infty \tilde\tau C(\tilde\tau)\,d\tilde\tau,
\]
we obtain
\begin{equation}
  \langle\Delta\tilde{\mathbf r}^2(\tilde t)\rangle
\approx
\frac{4\,\tilde t^{\alpha_T}}{\mathrm{Pe}^2\Gamma(1+\alpha_T)}
+2D_{\rm act}\tilde t-2B,
\qquad \tilde t\gg1.  
\end{equation}
Thus the active contribution becomes asymptotically diffusive, while the translational memory contributes a subleading subdiffusive term proportional to $\tilde t^{\alpha_T}$. The MSD exhibits two crossover times separating three distinct regimes:
\[
\langle\Delta\tilde{\mathbf r}^2(\tilde t)\rangle
\sim
\tilde t^{\alpha_T}
\quad\to\quad
\tilde t^2
\quad\to\quad
\tilde t.
\]
The first crossover, from translational-memory-dominated subdiffusion to ballistic active motion, is estimated by balancing the first two leading short-time terms, which gives
\begin{equation}
\tilde t_{c1}
\sim
\left[
\frac{4}{\mathrm{Pe}^2\Gamma(1+\alpha_T)}
\right]^{\frac{1}{2-\alpha_T}}.
\end{equation}

The second crossover, from ballistic motion to long-time diffusion, is estimated by balancing the ballistic contribution with the asymptotic diffusive term, $\tilde t^2\sim 2D_{\rm act}\tilde t$, yielding $\tilde t_{c2}\sim 2D_{\rm act}$,
where
\begin{equation}
D_{\rm act}=
\int_0^\infty
\cos(\tilde\omega_0\tau)
\exp\!\left[-\frac{\tau^{\alpha_R}}{\Gamma(1+\alpha_R)}\right]d\tau.
\end{equation}
In the nonchiral case $(\tilde\omega_0=0)$, this reduces to
\begin{equation}
\tilde t_{c2}
\sim
\frac{2}{\alpha_R}\,
\Gamma(1+\alpha_R)^{1/\alpha_R}
\Gamma\!\left(\frac{1}{\alpha_R}\right).
\end{equation}
\end{widetext}

\bibliography{revision1_ref_pwr_abm}

@article{Solon2015,
  author = {Solon, A. P. and Fily, Y. and Baskaran, A. and Cates, M. E. and Kafri, Y. and Kardar, M. and Tailleur, J.},
  title = {Pressure is not a state function for generic active fluids},
  journal = {Nat. Phys.},
  year = {2015},
  volume = {11},
  number = {8},
  pages = {673},
  doi = {10.1038/nphys3377}
}

@article{Stenhammar2014,
  title={Phase behaviour of active Brownian particles: the role of dimensionality},
  author={Stenhammar, J. and Marenduzzo, D. and Allen, R. J. and Cates, M. E.},
  journal={Soft Matter},
  volume={10},
  number={10},
  pages={1489},
  year={2014},
  publisher={Royal Society of Chemistry},
  doi = {10.1039/C3SM52813H}
}

@article{Redner2013,
  author = {Redner, G. S. and Hagan, M. F. and Baskaran, A.},
  title = {Structure and Dynamics of a Phase-Separating Active Colloidal Fluid},
  journal = {Phys. Rev. Lett.},
  year = {2013},
  volume = {110},
  number = {5},
  pages = {055701},
  doi = {10.1103/PhysRevLett.110.055701}
}

@article{Farage2015,
  author = {Farage, T. F. F. and Krinninger, P. and Brader, J. M.},
  title = {Effective interactions in active Brownian suspensions},
  journal = {Phys. Rev. E},
  year = {2015},
  volume = {91},
  number = {4},
  pages = {042310},
  doi = {10.1103/PhysRevE.91.042310}
}

@article{Berthier2014,
  title={Nonequilibrium glassy dynamics of self-propelled hard disks},
  author={Berthier, L.},
  journal={Phys. Rev. Lett.},
  volume={112},
  number={22},
  pages={220602},
  year={2014},
  publisher={APS},
  doi = {10.1103/PhysRevLett.112.220602}
}

@article{Turci2021wetting,
  title={Wetting transition of active Brownian particles on a thin membrane},
  author={Turci, F. and Wilding, N. B.},
  journal={Phys. Rev. Lett.},
  volume={127},
  number={23},
  pages={238002},
  year={2021},
  publisher={APS},
  doi={10.1103/PhysRevLett.127.238002}
}

@article{Adelman1976,
  author    = {Adelman, S. A.},
  title     = {Fokker–Planck equations for simple non‐Markovian systems},
  journal   = {J. Chem. Phys.},
  volume    = {64},
  number    = {1},
  pages     = {124},
  year      = {1976},
  doi       = {10.1063/1.431961}
}

@article{Angelini2011,
  author    = {Angelini, T. E. and Hannezo, E. and Trepat, X. and Marquez, M. and Fredberg, J. J. and Weitz, D. A.},
  title     = {Glass-like dynamics of collective cell migration},
  journal   = {Proc. Natl. Acad. Sci.},
  volume    = {108},
  number    = {12},
  pages     = {4714},
  year      = {2011},
  doi       = {10.1073/pnas.1010059108}
}

@article{Ariel2015,
  author    = {Ariel, G. and Rabani, A. and Benisty, S. and Partridge, J. D. and Harshey, R. M. and Be'er, A.},
  title     = {Swarming bacteria migrate by Levy walks},
  journal   = {Nat. Commun.},
  volume    = {6},
  number    = {1},
  pages     = {8396},
  year      = {2015},
  doi       = {10.1038/ncomms9396}
}

@article{Barkai2012,
  author    = {Barkai, E. and Garini, Y. and Metzler, R.},
  title     = {Strange kinetics of single molecules in living cells},
  journal   = {Phys. Today},
  volume    = {65},
  number    = {8},
  pages     = {29},
  year      = {2012},
  doi       = {10.1063/PT.3.1677}
}

@article{Bechinger2016,
  author    = {Bechinger, C. and Di Leonardo, R. and L{\"o}wen, H. and Reichhardt, C. and Volpe, G. and Volpe, G.},
  title     = {Active particles in complex and crowded environments},
  journal   = {Rev. Mod. Phys.},
  volume    = {88},
  number    = {4},
  pages     = {045006},
  year      = {2016},
  doi       = {10.1103/RevModPhys.88.045006}
}

@article{Bialke2013,
  author    = {Bialk{\'e}, J. and L{\"o}wen, H. and Speck, T.},
  title     = {Microscopic theory for the phase separation of self-propelled repulsive disks},
  journal   = {EPL},
  volume    = {103},
  number    = {3},
  pages     = {30008},
  year      = {2013},
  doi       = {10.1209/0295-5075/103/30008}
}

@article{Bouchaud1990,
  author    = {Bouchaud, J. P.},
  title     = {Anomalous diffusion in disordered media: Statistical mechanisms, models and physical applications},
  journal   = {Phys. Rep.},
  volume    = {195},
  number    = {4-5},
  pages     = {127},
  year      = {1990},
  doi       = {10.1016/0370-1573(90)90099-N}
}

@article{Bricard2013,
  author    = {Bricard, A. and Caussin, J.-B. and Desreumaux, N. and Dauchot, O. and Bartolo, D.},
  title     = {Emergence of macroscopic directed motion in populations of motile colloids},
  journal   = {Nature},
  volume    = {503},
  number    = {7474},
  pages     = {95},
  year      = {2013},
  doi       = {10.1038/nature12673}
}

@article{Burov2010,
  author    = {Burov, S. and Barkai, E.},
  title     = {Fractional Langevin Equation: Overdamped, Underdamped, and Critical Behaviors},
  journal   = {Phys. Rev. Lett.},
  volume    = {78},
  number    = {3},
  pages     = {031112},
  year      = {2008},
  doi       = {10.1103/PhysRevE.78.031112}
}

@article{Cates2015,
  author    = {Cates, M. E. and Tailleur, J.},
  title     = {Motility-Induced Phase Separation},
  journal   = {Annu. Rev. Condens. Matter Phys.},
  volume    = {6},
  pages     = {219},
  year      = {2015},
  doi       = {10.1146/annurev-conmatphys-031214-014710}
}

@article{Chen2007,
  author    = {Chen, D. T. and Wen, Q. and Janmey, P. A. and Crocker, J. C. and Yodh, A. G.},
  title     = {Rheology of Soft Materials},
  journal   = {Annu. Rev. Condens. Matter Phys.},
  volume    = {1},
  pages     = {301},
  year      = {2010},
  doi       = {10.1146/annurev-conmatphys-070909-104120}
}

@article{Cone2004,
  author    = {Cone, R. A.},
  title     = {Barrier properties of mucus},
  journal   = {Adv. Drug Deliv. Rev.},
  volume    = {61},
  number    = {2},
  pages     = {75},
  year      = {2009},
  doi       = {10.1016/j.addr.2008.09.008}
}

@inbook{Coffey2012,
  author    = {Coffey, W. T. and Kalmykov, Y. P.},
  title     = {The Langevin Equation: With Applications to Stochastic Problems in Physics, Chemistry and Electrical Engineering},
  publisher = {World Scientific},
  volume={27},
  year = {2012},
  doi={10.1142/8195}
}

@article{Crocker2000,
  author    = {Crocker, J. C. and Valentine, M. T. and Weeks, E. R. and Gisler, T. and Kaplan, P. D. and Yodh, A. G. and Weitz, D. A.},
  title     = {Two-Point Microrheology of Inhomogeneous Soft Materials},
  journal   = {Phys. Rev. Lett.},
  volume    = {85},
  number    = {4},
  pages     = {888},
  year      = {2000},
  doi       = {10.1103/PhysRevLett.85.888}
}

@article{Das2018,
  author    = {Das, S. and Gompper, G. and Winkler, R. G.},
  title     = {Confined active Brownian particles: theoretical description of propulsion-induced accumulation},
  journal   = {New J. Phys.},
  volume    = {20},
  number    = {1},
  pages     = {015001},
  year      = {2018},
  doi       = {10.1088/1367-2630/aa9d4b}
}

@article{ouyang2023swimming,
  title={Swimming of an inertial squirmer and squirmer dumbbell through a viscoelastic fluid},
  author={Ouyang, Z. and Lin, Z. and Lin, J. and Phan-Thien, N. and Zhu, J.},
  journal={J. Fluid Mech.},
  volume={969},
  pages={A34},
  year={2023},
  doi = {10.1017/jfm.2023.593}
}

@inbook{Doi1988,
  author    = {Doi, M. and Edwards, S. F.},
  title     = {The Theory of Polymer Dynamics},
  publisher = {Oxford University Press},
  year = {1988}
}

@article{Ebbens2016,
  author    = {Ebbens, S. J.},
  title     = {Active colloids: progress and challenges towards realising autonomous applications},
  journal   = {Curr. Opin. Colloid Interface Sci.},
  volume    = {21},
  pages     = {14},
  year      = {2016},
  doi       = {10.1016/j.cocis.2015.10.003}
}

@article{Elgeti2015,
  author    = {Elgeti, J. and Winkler, R. G. and Gompper, G.},
  title     = {Physics of microswimmers--single particle motion and collective behavior: a review},
  journal   = {Rep. Prog. Phys.},
  volume    = {78},
  number    = {5},
  pages     = {056601},
  year      = {2015},
  doi       = {10.1088/0034-4885/78/5/056601}
}

@article{Ahana2019confinement,
  title={Confinement induced trajectory of a squirmer in a two dimensional channel},
  author={Ahana, P. and Thampi, S. P.},
  journal={Fluid Dyn. Res.},
  volume={51},
  number={6},
  pages={065504},
  year={2019},
  publisher={IOP Publishing},
  doi={10.1088/1873-7005/ab4d08}
}

@article{Fily2012,
  author    = {Fily, Y. and Marchetti, M. C.},
  title     = {Athermal Phase Separation of Self-Propelled Particles with No Alignment},
  journal   = {Phys. Rev. Lett.},
  volume    = {108},
  number    = {23},
  pages     = {235702},
  year      = {2012},
  doi       = {10.1103/PhysRevLett.108.235702}
}

@article{Fox1977,
  author    = {Fox, R. F.},
  title     = {The generalized Langevin equation with Gaussian fluctuations},
  journal   = {J. Math. Phys.},
  volume    = {18},
  number    = {12},
  pages     = {2331},
  year      = {1977},
  doi       = {10.1063/1.523242}
}

@article{Garcia2015,
  author    = {Garcia, S. and Hannezo, E. and Elgeti, J. and Joanny, J.-F. and Silberzan, P. and Gov, N. S.},
  title     = {Physics of active jamming during collective cellular motion in a monolayer},
  journal   = {Proc. Natl. Acad. Sci.},
  volume    = {112},
  number    = {50},
  pages     = {15314},
  year      = {2015},
  doi       = {10.1073/pnas.1510973112}
}

@article{Mckinley2018anomalous,
  title={Anomalous diffusion and the generalized Langevin equation},
  author={McKinley, S. A. and Nguyen, H. D.},
  journal={SIAM J. Math. Anal.},
  volume={50},
  number={5},
  pages={5119},
  year={2018},
  publisher={SIAM},
  doi={10.1137/17M115517X}
}

@article{Ghosh2015,
  author    = {Ghosh, P. K. and Li, Y. and Marchegiani, G. and Marchesoni, F.},
  title     = {Communication: Memory effects and active Brownian diffusion},
  journal   = {J. Chem. Phys.},
  volume    = {143},
  number    = {21},
  pages     = {211101},
  year      = {2015},
  doi       = {10.1063/1.4936624}
}

@article{Gomez-Solano2016,
  author    = {Gomez-Solano, J. R. and Blokhuis, A. and Bechinger, C.},
  title     = {Dynamics of a Self-Propelled Particle in a Viscoelastic Fluid},
  journal   = {Phys. Rev. Lett.},
  volume    = {116},
  number    = {13},
  pages     = {138301},
  year      = {2016},
  doi       = {10.1103/PhysRevLett.116.138301}
}

@article{Gompper2020,
  author    = {Gompper, G. and Winkler, R. G. and Speck, T. and Solon, A. and Nardini, C. and Peruani, F. and L{\"o}wen, H. and Golestanian, R. and Kaupp, U. B. and Alvarez, L. and others},
  title     = {The 2020 motile active matter roadmap},
  journal   = {J. Phys. Condens. Matter},
  volume    = {32},
  number    = {19},
  pages     = {193001},
  year      = {2020},
  doi       = {10.1088/1361-648X/ab6348}
}

@article{Goychuk2012,
  author    = {Goychuk, I.},
  title     = {Viscoelastic Subdiffusion: Generalized Langevin Equation Approach},
  journal   = {Adv. Chem. Phys.},
  volume    = {150},
  pages     = {187},
  year      = {2012},
  doi       = {10.1002/9781118197714.ch5}
}

@article{Granick1995soft,
  title={Soft matter in a tight spot: nanorheology of confined liquids and block copolymers},
  author={Granick, S. and Demirel, A. L. and Cai, L. L. and Peanasky, J.},
  journal={Isr. J. Chem.},
  volume={35},
  number={1},
  pages={75},
  year={1995},
  publisher={Wiley Online Library},
  doi       = {10.1002/ijch.199500013}
}

@article{Guigas2007,
  author    = {Guigas, G. and Kalla, C. and Weiss, M.},
  title     = {Probing the nanoscale viscoelasticity of intracellular fluids in living cells},
  journal   = {Biophys. J.},
  volume    = {93},
  number    = {1},
  pages     = {316},
  year      = {2007},
  doi       = {10.1529/biophysj.106.099267}
}

@article{Hanggi1990,
  author    = {H{\"a}nggi, P. and Talkner, P. and Borkovec, M.},
  title     = {Reaction-rate theory: fifty years after Kramers},
  journal   = {Rev. Mod. Phys.},
  volume    = {62},
  number    = {2},
  pages     = {251},
  year      = {1990},
  doi       = {10.1103/RevModPhys.62.251}
}

@article{Hofling2013,
  author    = {H{\"o}fling, F. and Franosch, T.},
  title     = {Anomalous transport in the crowded world of biological cells},
  journal   = {Rep. Prog. Phys.},
  volume    = {76},
  number    = {4},
  pages     = {046602},
  year      = {2013},
  doi       = {10.1088/0034-4885/76/4/046602}
}

@article{Howse2007,
  author    = {Howse, J. R. and Jones, R. A. L. and Ryan, A. J. and Gough, T. and Vafabakhsh, R. and Golestanian, R.},
  title     = {Self-Motile Colloidal Particles: From Directed Propulsion to Random Walk},
  journal   = {Phys. Rev. Lett.},
  volume    = {99},
  number    = {4},
  pages     = {048102},
  year      = {2007},
  doi       = {10.1103/PhysRevLett.99.048102}
}

@article{Jeon2011,
  author    = {Jeon, J. H. and Metzler, R.},
  title     = {Fractional Brownian motion and motion governed by the fractional Langevin equation in confined geometries},
  journal   = {Phys. Rev. E},
  volume    = {81},
  number    = {2},
  pages     = {021103},
  year      = {2010},
  doi       = {10.1103/PhysRevE.81.021103}
}

@article{Kou2008,
  author    = {Kou, S. C.},
  title     = {Stochastic modeling in nanoscale biophysics: Subdiffusion within proteins},
  journal   = {Ann. Appl. Stat.},
  volume    = {2},
  number    = {2},
  pages     = {501},
  year      = {2008},
  doi       = {10.1214/07-AOAS149}
}

@article{Kubo1966,
  author    = {Kubo, R.},
  title     = {The fluctuation-dissipation theorem},
  journal   = {Rep. Prog. Phys.},
  volume    = {29},
  number    = {1},
  pages     = {255},
  year      = {1966},
  doi       = {10.1088/0034-4885/29/1/306}
}

@inbook{Larson1999,
  author    = {Larson, R. G.},
  title     = {The Structure and Rheology of Complex Fluids},
  publisher = {Oxford University Press},
  year      = {1999}
}

@article{Lauga2007,
  author    = {Lauga, E. and Powers, T. R.},
  title     = {The hydrodynamics of swimming microorganisms},
  journal   = {Rep. Prog. Phys.},
  volume    = {72},
  number    = {9},
  pages     = {096601},
  year      = {2009},
  doi       = {10.1088/0034-4885/72/9/096601}
}

@article{Patteson2016active,
  title={Active colloids in complex fluids},
  author={Patteson, A. E. and Gopinath, A. and Arratia, P. E.},
  journal={Curr. Opin. Colloid Interface Sci.},
  volume={21},
  pages={86},
  year={2016},
  publisher={Elsevier},
  doi={10.1016/j.cocis.2016.01.001}
}

@article{Loverdo2008,
  author    = {Loverdo, C. and B{\'e}nichou, O. and Moreau, M. and Voituriez, R.},
  title     = {Enhanced reaction kinetics in biological cells},
  journal   = {Nat. Phys.},
  volume    = {4},
  number    = {2},
  pages     = {134},
  year      = {2008},
  doi       = {10.1038/nphys830}
}

@article{Lowen2020,
  author    = {L{\"o}wen, H.},
  title     = {Inertial effects of self-propelled particles: From active Brownian to active Langevin motion},
  journal   = {J. Chem. Phys.},
  volume    = {152},
  number    = {4},
  pages     = {040901},
  year      = {2020},
  doi       = {10.1063/1.5134455}
}

@article{Lutz2001,
  author    = {Lutz, E.},
  title     = {Fractional Langevin equation},
  journal   = {Phys. Rev. E},
  volume    = {64},
  number    = {5},
  pages     = {051106},
  year      = {2001},
  doi       = {10.1103/PhysRevE.64.051106}
}

@inbook{Mainardi2010,
  author    = {Mainardi, F.},
  title     = {Fractional Calculus and Waves in Linear Viscoelasticity},
  publisher = {Imperial College Press},
  year      = {2010}
}

@article{Marchetti2013,
  author    = {Marchetti, M. C. and Joanny, J. F. and Ramaswamy, S. and Liverpool, T. B. and Prost, J. and Rao, Madan and Simha, R. Aditi},
  title     = {Hydrodynamics of soft active matter},
  journal   = {Rev. Mod. Phys.},
  volume    = {85},
  number    = {3},
  pages     = {1143},
  year      = {2013},
  doi       = {10.1103/RevModPhys.85.1143}
}

@article{Marconi2008,
  author    = {Marconi, U. M. B. and Puglisi, A. and Rondoni, L. and Vulpiani, A.},
  title     = {Fluctuation--dissipation: response theory in statistical physics},
  journal   = {Phys. Rep.},
  volume    = {461},
  number    = {4-6},
  pages     = {111},
  year      = {2008},
  doi       = {10.1016/j.physrep.2008.02.002}
}

@article{Mason1995,
  author    = {Mason, T. G. and Weitz, D. A.},
  title     = {Optical Measurements of Frequency-Dependent Linear Viscoelastic Moduli of Complex Fluids},
  journal   = {Phys. Rev. Lett.},
  volume    = {74},
  number    = {7},
  pages     = {1250},
  year      = {1995},
  doi       = {10.1103/PhysRevLett.74.1250}
}

@article{Metzler2000,
  author    = {Metzler, R. and Klafter, J.},
  title     = {The random walk's guide to anomalous diffusion: a fractional dynamics approach},
  journal   = {Phys. Rep.},
  volume    = {339},
  number    = {1},
  pages     = {1},
  year      = {2000},
  doi       = {10.1016/S0370-1573(00)00070-3}
}

@article{Metzler2014,
  author    = {Metzler, R. and Jeon, J.-H. and Cherstvy, A. G. and Barkai, E.},
  title     = {Anomalous diffusion models and their properties: non-stationarity, non-ergodicity, and ageing at the centenary of single particle tracking},
  journal   = {Phys. Chem. Chem. Phys.},
  volume    = {16},
  number    = {44},
  pages     = {24128},
  year      = {2014},
  doi       = {10.1039/C4CP03465A}
}

@article{Narinder2018,
  author    = {Narinder, N. and Bechinger, C. and Gomez-Solano, J. R.},
  title     = {Memory-Induced Transition from a Persistent Random Walk to Circular Motion for Achiral Microswimmers},
  journal   = {Phys. Rev. Lett.},
  volume    = {121},
  number    = {7},
  pages     = {078003},
  year      = {2018},
  doi       = {10.1103/PhysRevLett.121.078003}
}

@article{Palacci2013,
  author    = {Palacci, J. and Sacanna, S. and Steinberg, A. P. and Pine, D. J. and Chaikin, P. M.},
  title     = {Living crystals of light-activated colloidal surfers},
  journal   = {Science},
  volume    = {339},
  number    = {6122},
  pages     = {936},
  year      = {2013},
  doi       = {10.1126/science.1230020}
}

@article{Patteson2015,
  author    = {Patteson, A. E. and Gopinath, A. and Goulian, M. and Arratia, P. E.},
  title     = {Running and tumbling with E. coli in polymeric solutions},
  journal   = {Sci. Rep.},
  volume    = {5},
  number    = {1},
  pages     = {15761},
  year      = {2015},
  doi       = {10.1038/srep15761}
}

@article{Puertas2014,
  author    = {Puertas, A. M. and Voigtmann, Th.},
  title     = {Microrheology of colloidal systems},
  journal   = {J. Phys. Condens. Matter},
  volume    = {26},
  number    = {24},
  pages     = {243101},
  year      = {2014},
  doi       = {10.1088/0953-8984/26/24/243101}
}

@article{Ramaswamy2010,
  author    = {Ramaswamy, S.},
  title     = {The Mechanics and Statistics of Active Matter},
  journal   = {Annu. Rev. Condens. Matter Phys.},
  volume    = {1},
  pages     = {323},
  year      = {2010},
  doi       = {10.1146/annurev-conmatphys-070909-104101}
}

@article{Hallmann2015regulation,
  title={The regulation of immune cell trafficking by the extracellular matrix},
  author={Hallmann, R. and Zhang, X. and Di Russo, J. and Li, L. and Song, J. and Hannocks, M. J. and Sorokin, L.},
  journal={Curr. Opin. Cell Biol.},
  volume={36},
  pages={54},
  year={2015},
  publisher={Elsevier},
  doi={10.1016/j.ceb.2015.06.006}
}

@article{Romanczuk2012,
  author    = {Romanczuk, P. and B{\"a}r, M. and Ebeling, W. and Lindner, B. and Schimansky-Geier, L.},
  title     = {Active Brownian particles. From individual to collective statistics},
  journal   = {Eur. Phys. J.: Spec. Top.},
  volume    = {202},
  number    = {1},
  pages     = {1},
  year      = {2012},
  doi       = {10.1140/epjst/e2012-01529-y}
}

@article{Saintillan2018,
  author    = {Saintillan, D.},
  title     = {Rheology of active fluids},
  journal   = {Annu. Rev. Fluid Mech.},
  volume    = {50},
  pages     = {563},
  year      = {2018},
  doi       = {10.1146/annurev-fluid-010816-060049}
}

@inbook{Evangelista2018fractional,
  title={Fractional diffusion equations and anomalous diffusion},
  author={Evangelista, L. R. and Lenzi, E. K.},
  year={2018},
  publisher={Cambridge University Press}
}

@article{Luchko2016new,
  title={A new fractional calculus model for the two-dimensional anomalous diffusion and its analysis},
  author={Luchko, Y.},
  journal={Math. Model. Nat. Phenom.},
  volume={11},
  number={3},
  pages={1},
  year={2016},
  publisher={EDP Sciences},
  doi={10.1051/mmnp/201611301}
}

@article{Sandev2017,
  author    = {Sandev, T. and Metzler, R. and Chechkin, A.},
  title     = {From continuous time random walks to the generalized diffusion equation},
  journal   = {FCAA},
  volume    = {21},
  number    = {1},
  pages     = {10},
  year      = {2018},
  doi       = {10.1515/fca-2018-0002}
}

@article{Sevilla2019,
  author    = {Sevilla, F. J. and Rodriguez, R. F. and Gomez-Solano, J. R.},
  title     = {Active Ornstein-Uhlenbeck particle in a viscoelastic bath},
  journal   = {Phys. Rev. E},
  volume    = {100},
  number    = {3},
  pages     = {032123},
  year      = {2019},
  doi       = {10.1103/PhysRevE.100.032123}
}

@article{Sollich1997,
  author    = {Sollich, P. and Lequeux, F. and H{\'e}braud, P. and Cates, M. E.},
  title     = {Rheology of Soft Glassy Materials},
  journal   = {Phys. Rev. Lett.},
  volume    = {78},
  number    = {10},
  pages     = {2020},
  year      = {1997},
  doi       = {10.1103/PhysRevLett.78.2020}
}

@article{Sollich1998,
  author    = {Sollich, P.},
  title     = {Rheological constitutive equation for a model of soft glassy materials},
  journal   = {Phys. Rev. E},
  volume    = {58},
  number    = {1},
  pages     = {738},
  year      = {1998},
  doi       = {10.1103/PhysRevE.58.738}
}

@article{Spagnolie2012,
  author    = {Spagnolie, S. E. and Lauga, E.},
  title     = {Hydrodynamics of self-propulsion near a boundary: predictions and accuracy of far-field approximations},
  journal   = {J. Fluid Mech.},
  volume    = {700},
  pages     = {105},
  year      = {2012},
  doi       = {10.1017/jfm.2012.101}
}

@article{Sprenger2022,
  title = {Active Brownian motion with memory delay induced by a viscoelastic medium},
  author = {Sprenger, A. R. and Bair, C. and L\"owen, H.},
  journal = {Phys. Rev. E},
  volume = {105},
  issue = {4},
  pages = {044610},
  numpages = {8},
  year = {2022},
  month = {Apr},
  publisher = {American Physical Society},
  doi = {10.1103/PhysRevE.105.044610}
}

@article{Squires2010,
  author    = {Squires, T. M. and Mason, T. G.},
  title     = {Fluid mechanics of microrheology},
  journal   = {Annu. Rev. Fluid Mech.},
  volume    = {42},
  pages     = {413},
  year      = {2010},
  doi       = {10.1146/annurev-fluid-121108-145608}
}

@article{Tabei2013,
  author    = {Tabei, S. M. A. and Burov, S. and Kim, H. Y. and Kuznetsov, A. and Huynh, T. and Jureller, J. and Philipson, L. H. and Dinner, A. R. and Scherer, N. F.},
  title     = {Intracellular transport of insulin granules is a subordinated random walk},
  journal   = {Proc. Natl. Acad. Sci.},
  volume    = {110},
  number    = {13},
  pages     = {4911},
  year      = {2013},
  doi       = {10.1073/pnas.1221962110}
}

@article{tenHagen2011,
  author    = {ten Hagen, B. and van Teeffelen, S. and L{\"o}wen, H.},
  title     = {Brownian motion of a self-propelled particle},
  journal   = {J. Phys. Condens. Matter},
  volume    = {23},
  number    = {19},
  pages     = {194119},
  year      = {2011},
  doi       = {10.1088/0953-8984/23/19/194119}
}

@article{vicsek1995novel,
  title = {Novel Type of Phase Transition in a System of Self-Driven Particles},
  author = {Vicsek, T. and Czir\'ok, A. and Ben-Jacob, E. and Cohen, I. and Shochet, O.},
  journal = {Phys. Rev. Lett.},
  volume = {75},
  issue = {6},
  pages = {1226},
  numpages = {0},
  year = {1995},
  month = {Aug},
  publisher = {American Physical Society},
  doi={10.1103/PhysRevLett.75.1226}
}

@article{Vicsek2012,
  author    = {Vicsek, T. and Zafeiris, A.},
  title     = {Collective motion},
  journal   = {Phys. Rep.},
  volume    = {517},
  number    = {3-4},
  pages     = {71},
  year      = {2012},
  doi       = {10.1016/j.physrep.2012.03.004}
}

@article{chate2008modeling,
  title={Modeling collective motion: variations on the Vicsek model},
  author={Chat{\'e}, H. and Ginelli, F. and Gr{\'e}goire, G. and Peruani, F. and Raynaud, F.},
  journal={Eur. Phys. J B},
  volume={64},
  number={3},
  pages={451},
  year={2008},
  publisher={Springer},
  doi = {10.1140/epjb/e2008-00275-9}
}

@article{chate2008collective,
  title={Collective motion of self-propelled particles interacting without cohesion},
  author={Chat{\'e}, H. and Ginelli, F. and Gr{\'e}goire, G. and Raynaud, F.},
  journal={Phys. Rev. E},
  volume={77},
  number={4},
  pages={046113},
  year={2008},
  publisher={APS},
  doi = {10.1103/PhysRevE.77.046113}
}

@article{gregoire2004onset,
  title={Onset of collective and cohesive motion},
  author={Gr{\'e}goire, G. and Chat{\'e}, H.},
  journal={Phys. Rev. Lett.},
  volume={92},
  number={2},
  pages={025702},
  year={2004},
  publisher={APS},
  doi = {10.1103/PhysRevLett.92.025702}
}

@article{Waigh2005,
  author    = {Waigh, T. A.},
  title     = {Microrheology of complex fluids},
  journal   = {Rep. Prog. Phys.},
  volume    = {68},
  number    = {3},
  pages     = {685},
  year      = {2005},
  doi       = {10.1088/0034-4885/68/3/R04}
}

@article{Walther2013,
  author    = {Walther, A. and M{\"u}ller, A. H. E.},
  title     = {Janus Particles: Synthesis, Self-Assembly, Physical Properties, and Applications},
  journal   = {Chem. Rev.},
  volume    = {113},
  number    = {7},
  pages     = {5194},
  year      = {2013},
  doi       = {10.1021/cr300089t}
}

@article{Weber2012nonthermal,
  title={Nonthermal ATP-dependent fluctuations contribute to the in vivo motion of chromosomal loci},
  author={Weber, S. C. and Spakowitz, A. J. and Theriot, J. A.},
  journal={Proc. Natl. Acad. Sci.},
  volume={109},
  number={19},
  pages={7338},
  year={2012},
  publisher={National Academy of Sciences},
  doi={10.1073/pnas.1119505109}
}

@article{Weigel2011,
  author    = {Weigel, A. V. and Simon, B. and Tamkun, M. M. and Krapf, D.},
  title     = {Ergodic and nonergodic processes coexist in the plasma membrane as observed by single-molecule tracking},
  journal   = {Proc. Natl. Acad. Sci.},
  volume    = {108},
  number    = {16},
  pages     = {6438},
  year      = {2011},
  doi       = {10.1073/pnas.1016325108}
}

@article{Zhang2010,
  author    = {Zhang, H. P. and Be'er, A. and Florin, E.-L. and Swinney, H. L.},
  title     = {Collective motion and density fluctuations in bacterial colonies},
  journal   = {Proc. Natl. Acad. Sci.},
  volume    = {107},
  number    = {31},
  pages     = {13626},
  year      = {2010},
  doi       = {10.1073/pnas.1001651107}
}

@article{Zottl2019,
  author    = {Z{\"o}ttl, A. and Yeomans, J. M.},
  title     = {Enhanced bacterial swimming speeds in macromolecular polymer solutions},
  journal   = {Nat. Phys.},
  volume    = {15},
  number    = {6},
  pages     = {554},
  year      = {2019},
  doi       = {10.1038/s41567-019-0454-3}
}

@article{Gomez2020active,
  title={Active particles with fractional rotational Brownian motion},
  author={Gomez-Solano, J. R. and Sevilla, F. J.},
  journal={J. Stat. Mech.: Theory Exp.},
  volume={2020},
  number={6},
  pages={063213},
  year={2020},
  publisher={IOP Publishing},
  doi={10.1088/1742-5468/ab8553}
}

@inbook{Zwanzig2001,
  author    = {Zwanzig, R.},
  title     = {Nonequilibrium Statistical Mechanics},
  publisher = {Oxford University Press},
  year      = {2001}
}

@article{Zwanzig1973,
  author    = {Zwanzig, R.},
  title     = {Nonlinear generalized Langevin equations},
  journal   = {J. Stat. Phys.},
  volume    = {9},
  number    = {3},
  pages     = {215},
  year      = {1973},
  doi       = {10.1007/BF01008729}
}

@article{Theeyancheri2024,
  title = {Dynamic clustering of active rings},
  author = {Theeyancheri, L. and Chaki, S. and Bhattacharjee, T. and Chakrabarti, R.},
  journal = {Phys. Rev. Res.},
  volume = {6},
  issue = {1},
  pages = {L012038},
  numpages = {5},
  year = {2024},
  month = {Feb},
  publisher = {American Physical Society},
  doi = {10.1103/PhysRevResearch.6.L012038}
}

@article{di2011visco,
  title={Visco-elastic behavior through fractional calculus: an easier method for best fitting experimental results},
  author={Di Paola, M. and Pirrotta, A. and Valenza, A.},
  journal={Mechanics of materials},
  volume={43},
  number={12},
  pages={799},
  year={2011},
  publisher={Elsevier},
  doi={10.1016/j.mechmat.2011.08.016}
}

@article{bonfanti2020fractional,
  title={Fractional viscoelastic models for power-law materials},
  author={Bonfanti, A. and Kaplan, J. L. and Charras, G. and Kabla, A.},
  journal={Soft matter},
  volume={16},
  number={26},
  pages={6002},
  year={2020},
  publisher={Royal Society of Chemistry},
  doi={10.1039/d0sm00354a}
}

@article{Codina2022,
  title = {Small Obstacle in a Large Polar Flock},
  author = {Codina, J. and Mahault, B. and Chat\'e, H. and Dobnikar, J. and Pagonabarraga, I. and Shi, X. Q.},
  journal = {Phys. Rev. Lett.},
  volume = {128},
  issue = {21},
  pages = {218001},
  numpages = {6},
  year = {2022},
  month = {May},
  publisher = {American Physical Society},
  doi = {10.1103/PhysRevLett.128.218001}
}

@article{Netz2024,
  author  = {Netz, R. R.},
  title   = {Derivation of the nonequilibrium generalized Langevin equation from a time-dependent many-body Hamiltonian},
  journal = {Phys. Rev. E},
  volume  = {110},
  number  = {1},
  pages   = {014123},
  year    = {2024},
  doi     = {10.1103/PhysRevE.110.014123}
}

@article{Klimek2026,
  author  = {Klimek, A. and Baruah, P. V. and Sharma, P. and Netz, R. R.},
  title   = {Non-Markovian non-equilibrium modeling of experimental cell-motion trajectories reveals dependence of propulsion-force correlations on solvent viscosity},
  journal = {arXiv preprint arXiv:2601.18669},
  year    = {2026},
  doi     = {10.48550/arXiv.2601.18669}
}

@article{Quevedo2025,
  author  = {Quevedo, D. S. and Conte, M. and Dijkstra, M. and Morais Smith, C.},
  title   = {Active Brownian particles in power-law viscoelastic media},
  journal = {arXiv preprint arXiv:2512.20205},
  year    = {2025},
  doi     = {10.48550/arXiv.2512.20205}
}
\end{document}